\documentclass[aps,pra,superscriptaddress,twocolumn,nofootinbib,longbibliography]{revtex4-1}

\usepackage{amsfonts}
\usepackage{graphicx,graphics,epsfig,times,bm,bbm,amssymb,amsmath,amsfonts,mathrsfs}
\usepackage[normalem]{ulem}
\usepackage{wrapfig}
\usepackage{boxedminipage}
\usepackage{setspace}
\usepackage{subfigure}
\usepackage{dsfont}
\usepackage{braket}
\usepackage[pdftex]{color}
\usepackage[pdfstartview=FitH]{hyperref}
\usepackage{upgreek}
\usepackage{float}

\newcommand{\bes} {\begin{subequations}}
\newcommand{\ees} {\end{subequations}}
\newcommand{\bea} {\begin{eqnarray}}
\newcommand{\eea} {\end{eqnarray}}
\newcommand{\beq} {\begin{equation}}
\newcommand{\eeq} {\end{equation}}

\def\>{\rangle}
\def\<{\langle}
\def\Tr{\mathrm{Tr}}

\newcommand{\ketbra}[2]{|{#1}\>\<#2|}

\newcommand{\eps}{\varepsilon}
\newcommand{\ident}{\mathds{1}}
\newcommand{\ignore}[1]{}

\begin{document}
\title{Decoherence in adiabatic quantum computation}

\author{Tameem Albash}
\affiliation{Information Sciences Institute, University of Southern California, Marina del Rey, CA 90292}
\affiliation{Department of Physics and Astronomy, University of Southern California, Los Angeles, California
90089, USA}
\affiliation{Center for Quantum
  Information Science \& Technology, University of Southern California, Los Angeles, California
90089, USA}

\author{Daniel A. Lidar}
\affiliation{Department of Physics and Astronomy, University of Southern California, Los Angeles, California
90089, USA}
\affiliation{Center for Quantum
  Information Science \& Technology, University of Southern California, Los Angeles, California 90089, USA}
\affiliation{Department of Electrical Engineering, University of Southern California, Los Angeles, California
90089, USA}
\affiliation{Department of Chemistry, University of Southern California, Los Angeles, California
90089, USA}

\begin{abstract}
Recent experiments with increasingly larger numbers of qubits have sparked renewed interest in adiabatic quantum computation, and in particular quantum annealing.  A central question that is repeatedly asked is whether quantum features of the evolution can survive over the long time-scales used for quantum annealing relative to standard measures of the decoherence time.  We reconsider the role of decoherence in adiabatic quantum computation and quantum annealing using the adiabatic quantum master equation formalism.  We restrict ourselves to the weak-coupling and  singular-coupling limits, which correspond to decoherence in the energy eigenbasis and in the computational basis, respectively.  We demonstrate that  decoherence in the instantaneous energy eigenbasis does not necessarily detrimentally affect adiabatic quantum computation, and in particular that a short single-qubit $T_2$ time need not imply adverse consequences for the success of the quantum adiabatic algorithm. We further demonstrate that boundary cancellation methods, designed to improve the fidelity of adiabatic quantum computing in the closed system setting, remain beneficial in the open system setting. To address the high computational cost of master equation simulations, we also demonstrate that a quantum Monte Carlo algorithm that explicitly accounts for a thermal bosonic bath can be used to interpolate between classical and quantum annealing. Our study highlights and clarifies the significantly different role played by decoherence in the adiabatic and circuit models of quantum computing.
\end{abstract}

\maketitle
\section{Introduction}
%

Suppose qubits with dephasing time $T_2$ participate in a computation lasting a time $t_f$, without any error correction, yet $T_2 \ll t_f$. Can this computation be quantum? It is a commonly stated folklore position that the answer must be `no'. It is our goal in this paper to provide a theoretical basis for conditions under which the answer is in fact a qualified `yes'. 

Certainly, if one adopts the perspective of the circuit model of quantum computation, then the `no' answer is fully justified \cite{Aharonov:96a}. However, what is true in the circuit model does not necessarily apply directly to other models of quantum computation, in particular the adiabatic model \cite{FarhiAQC:00}, in spite of the fact that the two models are computationally equivalent \cite{aharonov_adiabatic_2007,PhysRevLett.99.070502,Gosset:2014rp}. Thus, a commonly held belief, that a short single-qubit dephasing time necessarily implies quantum computational failure, should not be applied without first carefully specifying the computational model.

We are motivated to revisit the question of the role of decoherence in adiabatic quantum computation (AQC) and quantum annealing (QA) \cite{childs_robustness_2001,PhysRevLett.95.250503,PhysRevA.72.042317,PhysRevA.71.032330,2002quant.ph.11152K,PhysRevA.79.022107,PhysRevA.80.022303,TAQC,ABLZ:12-SI,PhysRevA.75.062313,Vega:2010fk,PhysRevA.74.052330,PhysRevA.90.062120} by recent experiments involving increasingly larger numbers of niobium superconducting flux qubits using programmable quantum annealing devices built by D-Wave \cite{Dwave,Berkley:2010zr}. For such flux qubits, the $T_2$ time can at present range from tens of nanoseconds to a few hundred nanoseconds \cite{harris_flux_qubit_2010,kaiser_aluminum_2011}, yet the computation lasts on the order of microseconds to milliseconds. If the qubits have all decohered long before the computation is over, how can this be reconciled with evidence that the D-Wave devices perform quantum annealing \cite{DWave-16q,q-sig,q108,comment-SS,Crowley:2014qp,q-sig2,2014Katzgraber,DWave-entanglement,Boixo:2014yu,PAL:14}? In essence, the answer boils down to two key points: 
\begin{itemize}
\item The computation takes place in (or close to) the ground state, and 
\item Decoherence takes place in (or close to) the instantaneous energy eigenbasis. 
\end{itemize}
Using an adiabatic quantum master equation framework \cite{ABLZ:12-SI}, we shall explain how this leads to a very different behavior of adiabatic quantum computation in the presence of decoherence than what can be expected by direct analogy from the circuit model. The master equation provides a consistent framework for the analysis of decoherence in adiabatic quantum computation, which will help to clear up possible misconceptions arising from analogies drawn too closely with the circuit model. 

The structure of this paper is as follows. In Sec.~\ref{sec:II} we give a brief qualitative discussion of the role of the different timescales ($T_1$, $T_2$, and the total evolution time) in the circuit and adiabatic quantum computing models. In Sec.~\ref{sec:III} we summarize the adiabatic quantum master equation derived in Ref.~\cite{ABLZ:12-SI}, to set up the main tools used in this work. In particular, we distinguish between the weak and strong coupling limits (WCL and SCL, respectively), a distinction that gives rise to two very different master equations, and turns out to be crucial in understanding the role of decoherence in AQC. In Sec.~\ref{sec:IV} we apply these master equations to the simple case of a single qubit with a time-independent Hamiltonian, and re-derive familiar results in order to establish the appearance of the $T_1$ and $T_2$ times. We venture into new territory in Sec.~\ref{sec:dec-1q}, where we apply the WCL and SCL master equations to the case of a single-qubit coupled to a time-dependent Hamiltonian.  We consider both the adiabatic and non-adiabatic cases, and demonstrate explicitly that the role of decoherence is very different depending on whether the WCL or SCL applies. In particular, in the WCL thermally assisted AQC \cite{TAQC} can take place in the adiabatic limit, while in the SCL the population is distributed equally in the long time limit between the ground and excited state, so that AQC becomes impossible. We also demonstrate that the adiabatic limit is not optimal, in the sense that the ground state population is maximized at a total evolution time that can be much shorter than the adiabatic timescale and is determined by the bath spectral density. We digress in Sec.~\ref{sec:interp} to analyze the role of the interpolating function between the initial and final Hamiltonians on enhancing the probability of finding the ground state. We show that a strategy of imposing smooth boundary conditions developed for closed-system AQC in Ref.~\cite{lidar:102106} (see also \cite{Wiebe:12}) has a beneficial effect in the open-system setting as well, though the effect is milder. We then come to the analysis of AQC in the multi-qubit case, in Sec.~\ref{sec:VII}. Here is where we address the main question motivating this work, namely, the role of the single-qubit $T_2$ on the success probability of AQC. We again demonstrate that the answer depends drastically on whether the WCL or SCL applies, with the former's ground state population not exhibiting a dependence on the single-qubit $T_2$, under reasonable non-degeneracy assumptions. In Sec.~\ref{sec:VIII} we address a deficiency of our methodology, namely the fact that master equations are limited to a relatively small number of qubits (currently $\lesssim 15$), and the fact that there exists an intermediate coupling regime between the WCL and the SCL. We address this by considering a quantum Monte Carlo approach that explicitly accounts for a bosonic bath, and demonstrate its utility in interpolating between the quantum and classical regimes using a quantum annealing problem that has been studied in the context of quantumness tests of the D-Wave device. We conclude in Sec.~\ref{sec:IX}, and provide additional technical details in the Appendix.

\section{Timescales and decoherence in the circuit \textit{vs} the adiabatic model}
\label{sec:II}

The interaction between a quantum system and its environment is responsible for decoherence, which can undermine the efficiency of the quantum computation or even render it useless in the sense that it can be efficiently simulated by a classical computer \cite{Aharonov:96a}.  Two distinct decoherence time-scales are usually singled out.  The first is the loss of phase coherence between states, an elastic (energy conserving) process with a time-scale often referred to as $T_2$.  The second is the thermal equilibration time, which involves energy exchange with the thermal environment, with a time-scale often referred to as $T_1$.
In the simplest case of time-independent Markovian dynamics, where a system can be described by a Lindblad equation \cite{Lindblad:76,Gorini:1976uq,Alicki:87,Breuer:book}, there is a fundamental relation relating these two times scales, $T_2 \leq 2 T_1$ \cite{levitt2001spin}, but often it is the case that $T_2 \ll T_1$.  

In the quantum circuit paradigm, a quantum computation is a sequence of unitary operations (quantum gates) acting on one or more qubits. Quantum information is stored not only in the strings of $0$'s and $1$'s representing the state of the entire system in the computational basis, but also in the relative phase between superpositions of computational basis states. While models of circuit model quantum computation exist which allow for a high degree of decoherence and still enable a speedup over classical algorithms \cite{Knill:1998kx,Wu:2009}, it is clear that in general $T_2$ represents an upper limit on the time it takes to perform a circuit model quantum computation, in the absence of quantum error correction \cite{Lidar-Brun:book}.

In the adiabatic quantum computing paradigm \cite{FarhiAQC:00}, or in quantum annealing \cite{kadowaki_quantum_1998,PhysRevB.39.11828}, a computation is performed by evolving a system using a time-dependent Hamiltonian, with the final ground state encoding the solution of the computational problem. Thus the energy eigenbasis replaces the computational basis as the relevant basis for the computation, except possibly at the end, when the system may be measured in the computational basis, as is typical in adiabatic quantum optimization \cite{farhi_quantum_2001,RevModPhys.80.1061}. In this setting, 
phase coherence between energy eigenstates is irrelevant.  Moreover, if a near-optimal solution is acceptable then a computation which ends in a sufficiently low-lying excited state can be good enough and the condition that the computation terminates in the ground state can be relaxed \cite{King:2014uq} (though it follows from the PCP theorem that this does not necessarily change the complexity class \cite{Aharonov:2013vn}). Therefore, as long as the final measurement can clearly distinguish energy eigenstates then decoherence between energy states is harmless.  For this reason, unlike the circuit model, adiabatic quantum computation is believed to exhibit a degree of inherent robustness to decoherence \cite{childs_robustness_2001,PhysRevLett.95.250503,PhysRevA.72.042317,PhysRevA.71.032330,2002quant.ph.11152K}.

Let us now give a heuristic argument why the role of the total computation (i.e., evolution) time $t_f$ is quite different if one compares closed to open-system adiabatic quantum computation. In the closed-system setting, the only relevant time-scale is the condition that the evolution be sufficiently adiabatic, i.e., $T_{\textrm{ad}} \sim 1/\Delta_{\mathrm{min}}$, where $\Delta_{\mathrm{min}}$ is the minimum energy gap between the instantaneous ground state and all excited states that do not become part of the ground subspace. In the open-system setting, if phase decoherence does occur only in the energy eigenbasis, the remaining relevant time-scales determining the efficiency of the computation are $t_f$, the relaxation time $T_1$, and the time-scale associated with the (closed) system evolution being sufficiently adiabatic $T_{\mathrm{ad}}$. The interplay between these time-scales is non-monotonic and certainly more complicated than in the closed-system setting. In the latter, setting the heuristic adiabatic condition, $t_f \gg T_{\textrm{ad}}$, guarantees---by suppressing non-adiabatic transitions---that the final state reached has high overlap with the ground state of the final Hamiltonian \cite{Born:1928,JPSJ.5.435,Jansen:07,lidar:102106,Amin:09}. 
However, in the presence of a thermal bath, even if $t_f \gg 1/\Delta_{\mathrm{min}}$ there may still be significant loss of population from the ground state due to thermal processes.  The reason is that thermal excitation rates from the ground state typically grow as the gap shrinks and as $t_f$ grows, allowing the system to thermally relax into a Gibbs state that may have a significant population in low-lying energy eigenstates.  Such thermal relaxation can adversely impact the efficiency of the adiabatic quantum computation.  
Therefore, we find that the role of $t_f$, which was unambiguous in the closed-system case, becomes ambiguous when thermal processes are considered, and in general one can expect an optimal value of $t_f$ that is problem dependent \cite{Steffen:2003ys,PhysRevA.71.012331,ABLZ:12-SI,PhysRevLett.95.250503,crosson2014different}. Indeed, we shall demonstrate this in Sec.~\ref{sec:dec-1q} below for a specific model involving a qubit coupled to a bosonic bath with an Ohmic spectral density.

\section{Master Equations}
\label{sec:III}
%
The considerations above can be made rigorous by analyzing a system evolving in the presence of a thermal bath that is described in terms of an adiabatic master equation with time-dependent Lindblad operators \cite{ABLZ:12-SI}. Such a master equation description has the attractive feature that it guarantees the positivity of the density matrix at all times, but naturally requires certain assumptions and approximations.  

Consider a time-dependent system Hamiltonian 
\beq
H_S(t) \ket{\eps_a(t)} = \eps_a(t) \ket{\eps_a(t)}\ ,
\eeq
where the states $\{ \ket{\eps_a(t)} \}$ are the instantaneous energy eigenstates 
and the aforementioned gap is
\beq
\Delta_{\min} \equiv \min_{a,t}(\eps_a(t) - \eps_0(t))>0 ,
\eeq 
where $\ket{\eps_0(t)}$ is the instantaneous ground state and $\ket{\eps_a(t)}$ ($a\geq 1$) are the excited states. The condition $\Delta_{\mathrm{min}} >0$ ensures that only excited states that do not eventually become part of the ground subspace are considered. 

Next consider a generic system-bath Hamiltonian,
\bes
\label{eq:H_general}
\begin{align}
H(t) &= H_S(t) \otimes \ident_B  + \ident_S\otimes H_B + H_I\\
H_I &= g \sum_{\alpha} A_{\alpha} \otimes B_{\alpha} \ ,
\end{align}
\ees
where $A_{\alpha}$ and $B_{\alpha}$ in the interaction Hamiltonian are, respectively, dimensionless Hermitian system and bath operators and $g$ is the system-bath coupling strength. An adiabatic master equation in Lindblad form \cite{Lindblad:76} for the system's evolution can be derived in the weak coupling limit (WCL)---where $H_S$ dominates $H_{SB}$ in the sense of Eq.~\eqref{eq:8a} below---by invoking the standard Born-Markov and rotating wave approximations, along with an adiabatic approximation \cite{ABLZ:12-SI}. 

Consider the bath correlation functions (we set $\hbar = 1$ henceforth):
\beq
\mathcal{B}_{\alpha \beta}(t) \equiv e^{i H_B t} B_{\alpha} e^{-i H_B t} B_{\beta}\ .
\eeq
The characteristic decay time $\tau_B$ is then defined via
\beq
|\braket{\mathcal{B}_{\alpha \beta}(t)}| \equiv |\Tr[\rho_B \mathcal{B}_{\alpha \beta}(t)]| \sim e^{-t/\tau_B},
\label{eq:tauB}
\eeq 
where $\rho_B$ is the initial state of the bath. Note that this exponential decay is not guaranteed but simply assumed here in order to extract the timescale $\tau_B$.

Now assume:
\bes
\begin{align}
\label{eq:8a}
g^2\tau_B &\ll \Delta_{\min} \qquad \textrm{(weak coupling)}\\
\label{eq:8b}
g\tau_B &\ll 1 \qquad \textrm{(Markov approximation)}\\
\label{eq:8c}
\frac{h}{t_f} &\ll \min\{\Delta_{\min}^2,\tau_B^{-2}\} ,
\end{align}
\ees
where $h \equiv \max_{s\in[0,1];a,b}|\bra{\eps_a(s)}\partial_s H(s)\ket{\eps_b(s)}|$ estimates the rate of change of the Hamiltonian. Inequality \eqref{eq:8c} combines the heuristic adiabatic approximation with the condition that the instantaneous energy eigenbasis should be slowly varying on the timescale of the bath \cite{ABLZ:12-SI}. 

Provided these conditions are satisfied the quantum adiabatic master equation takes the generic form \cite{ABLZ:12-SI}:
\bes
\label{eqt:ME2}
\begin{align}
\label{eqt:ME2-H}
\frac{d}{dt} \rho(t) &= -i \left[H_S(t) + H_{LS}(t), \rho(t) \right] + \mathcal{L}_{\textrm{WCL}} [\rho(t)] \\
\label{eqt:ME2-L}
\mathcal{L}_{\textrm{WCL}}  [\rho(t)]  & \equiv \sum_{\omega} \gamma_{\alpha \beta}(\omega) \left(L_{\beta,\omega}(t) \rho(t) L_{\alpha, \omega}^{\dagger}(t)\phantom{\frac{1}{2}} \right. \notag \\
& \left. \qquad\qquad - \frac{1}{2} \left\{ L_{\alpha,\omega}^{\dagger}(t) L_{\beta,\omega}(t) , \rho(t) \right\} \right) \ ,
\end{align}
\ees
where the sum over $\omega$ is over the Bohr frequencies of $H_{S}$, and where the time-dependent Lindblad operators are
\beq
L_{\alpha, \omega}(t)  = \sum_{\omega = \eps_b(t) - \eps_a(t)} \bra{\eps_a(t)} A_\alpha \ket{\eps_b(t)}\ketbra{\eps_a(t)}{\eps_b(t)} \ .
\label{eq:Lindblad2}
\eeq
The decay rates
\beq 
\gamma_{\alpha \beta}(\omega)  = g^2 \int_{-\infty}^{\infty} dt \ e^{i \omega t} \braket{\mathcal{B}_{\alpha \beta}(t)} 
\label{eq:gamma_ab}
\eeq
are Fourier transforms of the bath correlation function forming a positive matrix $\gamma(\omega)$ whose elements satisfy the KMS condition\footnote{We use a slightly different convention than Ref.~\cite{ABLZ:12-SI} by including the factor $g^2$ in the definitions of $\gamma_{\alpha \beta}(\omega)$ and $S_{\alpha \beta}(\omega)$ instead of pulling it out as in Eqs.~(46)-(50) in Ref.~\cite{ABLZ:12-SI}.}
\beq
\gamma_{\alpha \beta}(-\omega) = e^{-\upbeta \omega} \gamma_{\beta\alpha}(\omega)\ ,
\label{eq:KMS}
\eeq 
where $\upbeta$ is the inverse temperature, and 
\beq
H_{\textrm{LS}}  =  \sum_{\alpha \beta} \sum_{\omega} S_{\alpha \beta} (\omega) L_{\alpha,\omega}^{\dagger}(t) L_{\beta,\omega}(t) \ ,
\label{eq:H_LS}
\eeq
is a Lamb shift term, where 
\beq 
\label{eqt:ME2b}
S_{\alpha \beta}(\omega) = \int_{-\infty}^{\infty} d \omega' \gamma_{\alpha \beta}(\omega') \mathcal{P} \left( \frac{1}{\omega - \omega'} \right) \ ,
\eeq
with $\mathcal{P}$ denoting the Cauchy principal value.

The Lindblad master equation in Eq.~\eqref{eqt:ME2} can be thought of as a natural generalization of the time-independent case \cite{Breuer:book}, where the Hamiltonian varies sufficiently slowly relative to the bath so that at any instant in time we simply have a copy of the time-independent master equation.  In fact, we can recover the time-independent result by replacing $H_S(t)$ by a time-independent system Hamiltonian.  We will show in subsequent examples that, because of the form of the Lindblad operators in Eq.~\eqref{eq:Lindblad2},  \emph{decoherence occurs in the instantaneous energy eigenbasis}. 

Another standard case is the other extreme limit, where $H_{SB}$ dominates $H_{S}$ (i.e., where inequality \eqref{eq:8a} is reversed), which is often called the singular coupling limit (SCL).  The resulting master equation takes the form:
\bes
\label{eqt:SCL}
\begin{align}
\frac{d}{dt} \rho(t) &= -i \left[H_S(t) + H_{\textrm{LS}}, \rho(t) \right]  + \mathcal{L}_{\textrm{SCL}} [\rho(t)]\\
\mathcal{L}_{\textrm{SCL}} [\rho(t)] &\equiv \sum_{\alpha , \beta}   \gamma_{\alpha \beta}(0) \left(A_{\beta} \rho(t)A_{\alpha}^{\dagger} - \frac{1}{2} \left\{ A_{\alpha}^{\dagger} A_{\beta} , \rho(t) \right\} \right)  \ ,
\end{align}
\ees
where now $H_{\textrm{LS}} =  \sum_{\alpha \beta}  S_{\alpha \beta} (0) A_{\alpha}^{\dagger} A_{\beta}$.  In this limit, the Lindblad operators are simply the bare system operators $A_{\alpha}$, as is often written down in phenomenological treatments of the master equation.  Here \emph{decoherence occurs in the basis that diagonalizes these operators}, when such a basis exists, as we will show explicitly later.
%
\section{Decoherence of a single qubit with a time-independent Hamiltonian}
\label{sec:IV}
\subsection{Pure dephasing}
We begin our discussion with a quick review of the simplest case of decoherence.  We consider a single qubit with a time-independent Hamiltonian coupled to an arbitrary bath for which the conditions required for the derivation of the master equation are satisfied. 
The total Hamiltonian is given by Eq.~\eqref{eq:H_general}, but we shall assume that
\begin{align} 
\label{eq:Hqubit}
H_S = -\frac{1}{2}\omega_z \sigma^z \ , \qquad
H_I =  g\sigma^z \otimes B .
\end{align}
For the interaction Hamiltonian in Eq.~\eqref{eq:Hqubit}, there is only a single system operator $A  = \sigma^z = \ketbra{0}{0}-\ketbra{1}{1}$.  The eigenstates are $\ket{\eps_0(t)}=\ket{0}$ and $\ket{\eps_1(t)}=\ket{1}$. Considering the WCL master equation \eqref{eqt:ME2} and $\bra{\eps_a(t)} A \ket{\eps_b(t)} \propto \delta_{a,b}$, there is only a single Lindblad operator that is non-zero:
\beq 
\label{eqt:LindbladOp_Z}
L_{z,0} = \sigma^z\ ,
\eeq
as given by Eq.~\eqref{eq:Lindblad2}.  This follows since $[H_S,H_I]=0$.  Therefore, the master equation for the single qubit takes the simple form
\begin{align}
\frac{d}{d t} \rho(t) &= -i \left[ H_S, \rho(t) \right] \\
&+ \gamma(0) \left( L_{z,0} \rho(t) L_{z,0}^{\dagger} - \frac{1}{2} \left\{ L_{z,0}^{\dagger} L_{z,0} , \rho(t) \right\} \right) \notag ,
\end{align}
where we have also used the fact that $H_{\textrm{LS}} \propto \ident$.  This equation can be solved analytically and gives, after expanding $\rho(t) = \sum_{i,j\in\{0,1\}}\rho_{ij}\ketbra{i}{j}$, and taking matrix elements of Eq.~\eqref{eqt:ME2} (in the computational basis, which here is equivalent to the energy eigenbasis):
\bes
\begin{align}
\rho_{0 0}(t) & = \rho_{0 0}(0) = 1-  \rho_{1 1}(t)\ , \\
\rho_{0 1}(t) & = \exp(- t/T_2^{(c)} + i \omega_z t) \rho_{0 1}(0) = \rho_{1 0}^*(t) \ ,
\end{align}
\ees
%
%
%
where 
\beq 
\label{eqt:T2Z}
T_2^{(c)} = \frac{1}{2 \gamma(0)} \ ,
\eeq
where the `$c$' superscript denotes the computational basis (we shall shortly see a second $T_2$ associated with the energy eigenbasis). This is the familiar pure dephasing channel, where only the off-diagonals elements (transverse magnetization) decay with a characteristic timescale $T_2^{(c)}$. The stronger the coupling to the bath [recall via Eq.~\eqref{eq:gamma_ab} that $\gamma\propto g^2$], the shorter the qubit coherence time.  We note that the energy gap $\omega_z$ plays no role in the result for $T_2^{(c)}$, and in fact, $T_2^{(c)}$ here is entirely determined by the spectrum of the bath correlation function at zero frequency.  In this example there is no thermal relaxation (the $T_1$ time is infinite), since the population of the energy states remains fixed, as a consequence of $[H_S,H_I]=0$.

\subsection{Decoherence when $[H_S,H_I]\neq 0$}
Let us now replace the system Hamiltonian so that $[H_S,H_I]\neq 0$. Specifically, consider 
\begin{align} 
\label{eq:Hqubit2}
H_S = -\frac{1}{2}\omega_x \sigma^x\ , \qquad 
H_I =  g\sigma^z \otimes B\ .
\end{align}
We shall study this scenario in both the weak and singular coupling limits. We shall see that there is a sharp contrast between the two, with the WCL resulting in decoherence in the energy eigenbasis, while the SCL results in decoherence in the computational basis, just as in the previous subsection, when $H_S$ and $H_I$ were commuting.

\subsubsection{WCL}
\label{sec:WCL1}
The energy eigenstates of $H_S$ are $\ket{\eps_0} = \ket{+}$ and $\ket{\eps_1}~=~\ket{-}$ with respective eigenvalues $-\frac{1}{2} \omega_x$ and $\frac{1}{2} \omega_x$, where $\ket{\pm} = \frac{1}{\sqrt{2}} \left( \ket{0} \pm \ket{1} \right)$.  Since $\sigma^z \ket{\pm} =  \ket{\mp}$ the non-zero Lindblad operators are:
\beq \label{eqt:LindbladOp_X}
L_{z,\omega_x} = \ketbra{+}{-} \ , \quad L_{z,-\omega_x} = \ketbra{-}{+} \ .
\eeq
Note that we now have a non-trivial Lamb shift term:
\beq
H_{\textrm{LS}} = S(\omega_x) \ketbra{-}{-} + S(-\omega_x) \ketbra{+}{+} \ .
\eeq
Writing $\rho(t) = \sum_{i,j\in\{+,-\}}\rho_{ij}\ketbra{i}{j}$, and taking matrix elements of Eq.~\eqref{eqt:ME2}, we find that the master equation for the density matrix components is:
\bes 
\begin{align}
 \frac{d}{dt} \rho_{--}(t)  &= - \gamma(\omega_x) \rho_{--}(t) + \gamma(-\omega_x) \rho_{++}(t) \ , \\
\frac{d}{dt} \rho_{++}(t)  & = \gamma(\omega_x)  \rho_{--}(t) - \gamma(-\omega_x) \rho_{++}(t) \ , \\
\frac{d}{dt} \rho_{-+}(t)  & = \frac{d}{dt} \rho_{+-}^*(t)  = \left[ - i \left( S(\omega_x) - S(-\omega_x) + \omega_x \right) \phantom{\frac{1}{2}} \right. \notag \\ 
&\qquad  \left. 
- \frac{1}{2} \gamma(\omega_x) \left( 1 + e^{-\upbeta \omega_x} \right) \right] \rho_{-+}(t) \ ,
\end{align}
\ees
where we have used the KMS condition \eqref{eq:KMS} to simplify the expressions.  These equations can be solved analytically to give:
\bes \label{eqt:decoherence2}
\begin{align}
\rho_{-+}(t)  &= 2 \rho_{ - +}(0) e^{-i\left( S(\omega_x) - S(-\omega_x) + \omega_x \right) t} e^{-t / T_2^{(e)}} \ , \\
\rho_{--}(t) & = p_{\textrm{Gibbs}}(-) + \left[ \rho_{--}(0) -  p_{\textrm{Gibbs}}(-)  \right] e^{- t / T_1^{(e)}} \ , \\
\rho_{++}(t) & = 1- \rho_{--}(t) \ , \qquad \rho_{+-}(t)  = \rho_{-+}^*(t) \ ,
\end{align}
\ees
where
\beq
p_{\textrm{Gibbs}}(\pm) = \frac{e^{\pm\upbeta \omega_x/ 2}}{Z}\ , \qquad Z = e^{\beta \omega_x/2} + e^{-\upbeta \omega_x/2}\ ,
\eeq
and
\beq 
\label{eqt:T2X}
T_1^{(e)} = \frac{1}{\gamma(\omega_x) \left( 1 + e^{-\upbeta \omega_x} \right)} \ , \quad T_2^{(e)} = 2 T_1^{(e)} \ .
\eeq

We observe three important facts about the result in Eq.~\eqref{eqt:decoherence2}.  First, the decoherence occurs in the energy eigenbasis, i.e., the off-diagonal components \emph{in the energy eigenbasis} (hence the `$e$' superscripts on $T_1$ and $T_2$) decay exponentially to zero with a timescale determined by $T_2^{(e)}$, and this includes the entire contribution of the Lamb shift.  Second, the populations ($\rho_{++},\rho_{--}$) approach the Gibbs state associated with the Hamiltonian $H_S$ within a timescale determined by $T_1^{(e)}$.  Third, the two timescales ($T_1^{(e)},T_2^{(e)}$) have a non-trivial dependence on the energy gap $\omega_x$.

\subsubsection{SCL}
\label{sec:SCL1}
Let us contrast this with what happens in the SCL case, Eq.~\eqref{eqt:SCL}.  In this case the evolution of the off-diagonal components is most conveniently written in the computational basis:
\bes
\begin{align}
\frac{d}{dt} \rho_{0 0} & = -i \frac{1}{2} \omega_x \left(  \rho_{1 0} - \rho_{0 1} \right) \ , \\
\frac{d}{dt} \rho_{1 1} & = -i \frac{1}{2} \omega_x \left(  \rho_{0 1} -  \rho_{1 0} \right) \ , \\
\frac{d}{dt} \rho_{0 1} &=  i \frac{1}{2} \omega_x \left( \rho_{1 1} - \rho_{0 0} \right) - 2 \gamma(0) \rho_{0 1} \ , \\
\frac{d}{dt} \rho_{1 0} &=  i \frac{1}{2} \omega_x \left(\rho_{0 0} -  \rho_{1 1}  \right) - 2 \gamma(0) \rho_{1 0 } \ .
\end{align}
\ees
This set of equations can be solved analytically for arbitrary initial conditions, but for brevity, let us consider the case where the density matrix is initially in the ground state, i.e., $\rho(0) = \ketbra{+}{+}$.  The solution is then given by:
\beq \label{eqt:SCL_sol}
\rho_{0 0} = \rho_{1 1}= \frac{1}{2} \ , \quad \rho_{0 1} = \rho_{1 0} =  \frac{1}{2} e^{- t/T_2^{(c)} } \ .
\eeq
In this case, the off-diagonal elements \emph{in the computational basis} decay exponentially with a timescale determined by $T_2^{(c)}$ [Eq.~\eqref{eqt:T2Z}], so we have decoherence in the computational basis regardless of the fact that the system Hamiltonian does not commute with $H_I$.

These simple time-independent examples anticipate what we shall see for their time-dependent counterparts.  In the WCL, we expect decoherence in the energy eigenbasis, a feature that does not preclude the success of an adiabatic quantum computation, whereas in the SCL, we expect  decoherence in the computational basis, rendering adiabatic quantum computation impossible.
%
\section{Decoherence of a single qubit with a time-dependent Hamiltonian}
\label{sec:dec-1q}
We consider the following 
time-dependent single qubit Hamiltonian, with a linear interpolation schedule:
\beq
H_S(t) = - \frac{1}{2}{\omega_x} \left( 1 -s \right) \sigma^x - \frac{1}{2} s \omega_z \sigma^z \ ,
\label{eq:H_S-lin(t)}
\eeq
where $t\in[0,t_f]$ and $s = t / t_f$ is the dimensionless time (the role of the interpolation schedule is studied in Sec.~\ref{sec:interp}). This Hamiltonian interpolates between the two system Hamiltonians considered in Eqs.~\eqref{eq:Hqubit} and \eqref{eq:Hqubit2} in the time-independent case above. The instantaneous energy eigenvalues are:
\beq \label{eqt:1qubitenergies}
\eps_\pm(s) = \pm \frac{1}{2} \sqrt{ (1-s)^2{\omega_x^2} + s^2 \omega_z^2 } 
\equiv \pm \frac{\Delta(s)}{2} \ ,
\eeq
and hence the instantaneous energy gap is $\Delta(s)$. Its minimum is 
\beq
\Delta_{\min} = 
\frac{\omega_z \omega_x }{\sqrt{\omega_z^2+\omega_x^2}} \ ,
\label{eq:l-min}
\eeq
which is reached when $s=s_{\min}$, where
\beq
s_{\min}=\frac{1}{1+\Gamma^2}\ ; \qquad \Gamma \equiv \omega_z/\omega_x\ . 
\eeq
It is convenient to define the dimensionless instantaneous gap
\beq
\lambda(s) \equiv \frac{\Delta(s)}{\omega_x} = \sqrt{(1-s)^2+s^2 \Gamma^2}\ , 
\eeq
so that $\lambda_{\min} = \Delta_{\min} /\omega_x = {\Gamma }/\sqrt{1+\Gamma^2}$. 
The corresponding energy eigenstates are
\bes \label{eqt:1qubiteigenstates}
\begin{align}
\ket{\eps_+(s)} & = \frac{1}{c_+(s)} \left[ \frac{s \Gamma-\lambda(s)}{1 -s }  \ket{0} + \ket{1} \right] \ ,  \\
\ket{\eps_-(s)} & = \frac{1}{c_-(s)} \left[ (s \Gamma + \lambda(s)) \ket{0} + (1-s) \ket{1} \right] \ ,
\end{align}
\ees
with $c_{\pm}(s)$ being the appropriate normalization factors.
Note that when $s = 1$ the ground state is $\ket{0}$.

To identify the adiabatic limit for this case, let us use the heuristic adiabatic condition [Eq.~\eqref{eq:8c}]:
\beq
\max_{0 \leq s \leq 1} \frac{| \bra{\eps_+ } \partial_s H \ket{\eps_-}|}{t_f \Delta(s)^2}  = \frac{\frac{1}{2} \sqrt{1 + \Gamma^2} \omega_x}{t_f \Delta_{\min}^2} \ll 1 \ ,
\eeq
where we have used the fact that the numerator (which equals $\Gamma \omega_x /[2\lambda(s)]$ before maximization) and denominator are both respectively maximized and minimized at $s=s_{\min}$. Rewriting, this yields for the adiabatic condition
\bes
\begin{align}
\label{eq:36}
& t_f \omega_x \gg  \frac{\Gamma}{2\lambda_{\min}^3} \quad \textrm{or, equivalently} \\
& t_f\sqrt{\omega_x \omega_z} \gg \frac{1}{2} \left(\Gamma+\Gamma^{-1}\right)^{3/2} \ ,
\end{align}
 \ees
where the second form of the inequality emphasizes the symmetry between $\omega_x$ and $\omega_z$.


\subsection{WCL}
\label{sec:WCL2}
In the WCL, the Lindblad operators take the form:
\bes
\begin{align}
L_0(s) &= \frac{s\Gamma}{\lambda(s)} \left( \ketbra{\eps_-(s)}{\eps_-(s)} - \ketbra{\eps_+(s)}{\eps_+(s)} \right) \\  
L_{\pm\lambda}(s) &= \zeta(s) \ketbra{\eps_\mp(s)}{\eps_\pm(s)} \ ,
\end{align}
\ees
where
\beq
\zeta(s) \equiv \frac{1-s}{\lambda(s)}\ .
\eeq
(We have dropped the index $\alpha$ from the Lindblad operators since there is only a single qubit.)  Note that $0\leq \zeta(s) \leq 1$ and that these Lindblad operators interpolate between the operators in Eqs.~\eqref{eqt:LindbladOp_Z} and \eqref{eqt:LindbladOp_X}.  Denoting $\rho_{ij}(s) \equiv \bra{\eps_i(s)} \rho(s) \ket{\eps_j(s)}$ (with $i,j\in\{+,-\}$), and using $\bra{\eps_+} \partial_s \ket{\eps_-} = - \bra{\eps_-} \partial_s \ket{\eps_+}  = - \Gamma / 2 \lambda(s)^2$, we have as equations of motion:
\bes
\label{eq:32}
\begin{align}
\label{eq:32a}
\frac{d}{ds} & \rho_{--}(s)   =  -\frac{d}{ds} \rho_{++}(s) \notag \\ 
& = \frac{\Gamma}{2 \lambda^2(s)} \left( \rho_{-+} + \rho_{+-} \right) 
+ \left[  \mathcal{F}_{+}(s)  \rho_{++} - \mathcal{F}_{-}(s) \rho_{--} \right] \ , \\
\frac{d}{ds} &\rho_{+-}(s)  = \frac{d}{ds} \rho_{-+}^{\ast}(s) \notag \\
&= \frac{\Gamma}{2 \lambda^2(s)} \left( \rho_{++} - \rho_{--} \right) -  \big[ i \Omega(s) +   \Sigma(s)  \big]\rho_{+-} \ ,
 \label{eqt:1qubitPhase}
\end{align}
\ees
where
\bes
\begin{align}
\mathcal{F}_{\pm} (s) &=   t_f \zeta^2(s) \gamma(\pm \Delta(s)) \\ 
\Omega(s) &=   t_f \left[ \Delta(s) + \left\{ S(\Delta(s)) - S(-\Delta(s)) \right\} \zeta^2(s) \right] \\ 
\Sigma(s) &=   t_f \left[ 2 \gamma(0) \left(\frac{s \Gamma}{\lambda(s)} \right)^2 +  \right. \notag \\
& \qquad \qquad \left.  \frac{1}{2}\left[ \gamma(\Delta(s)) + \gamma(-\Delta(s)) \right] \zeta^2(s) \right]
\label{eqt:Sigma}
\end{align}
\ees
The  term of Eq.~\eqref{eqt:1qubitPhase} involving $\Sigma(s)$ gives rise to the exponential decay of the off-diagonal elements of the density matrix in the instantaneous energy eigenbasis. The corresponding decoherence time [$t_f/\Sigma(s)$ in Eq.~\eqref{eqt:Sigma}] interpolates between the $T_2^{(c)}$ and $T_2^{(e)}$ times given in Eqs.~\eqref{eqt:T2Z} and \eqref{eqt:T2X}, respectively:  $t_f/\Sigma(0) = 2/[\gamma(\omega_x) ( 1 + e^{-\upbeta \omega_x} )] = T_2^{(e)}$ [where we used the KMS condition \eqref{eq:KMS}] and $t_f/\Sigma(1) = 1/[2\gamma(0)] = T_2^{(c)}$. This is to be expected since as mentioned above, $H_S(t)$ [Eq.~\eqref{eq:H_S-lin(t)}] interpolates between the corresponding two system Hamiltonians.

\subsubsection{Solution in the adiabatic limit}
\begin{figure}[t] 
   \centering
   \includegraphics[width=0.9\columnwidth]{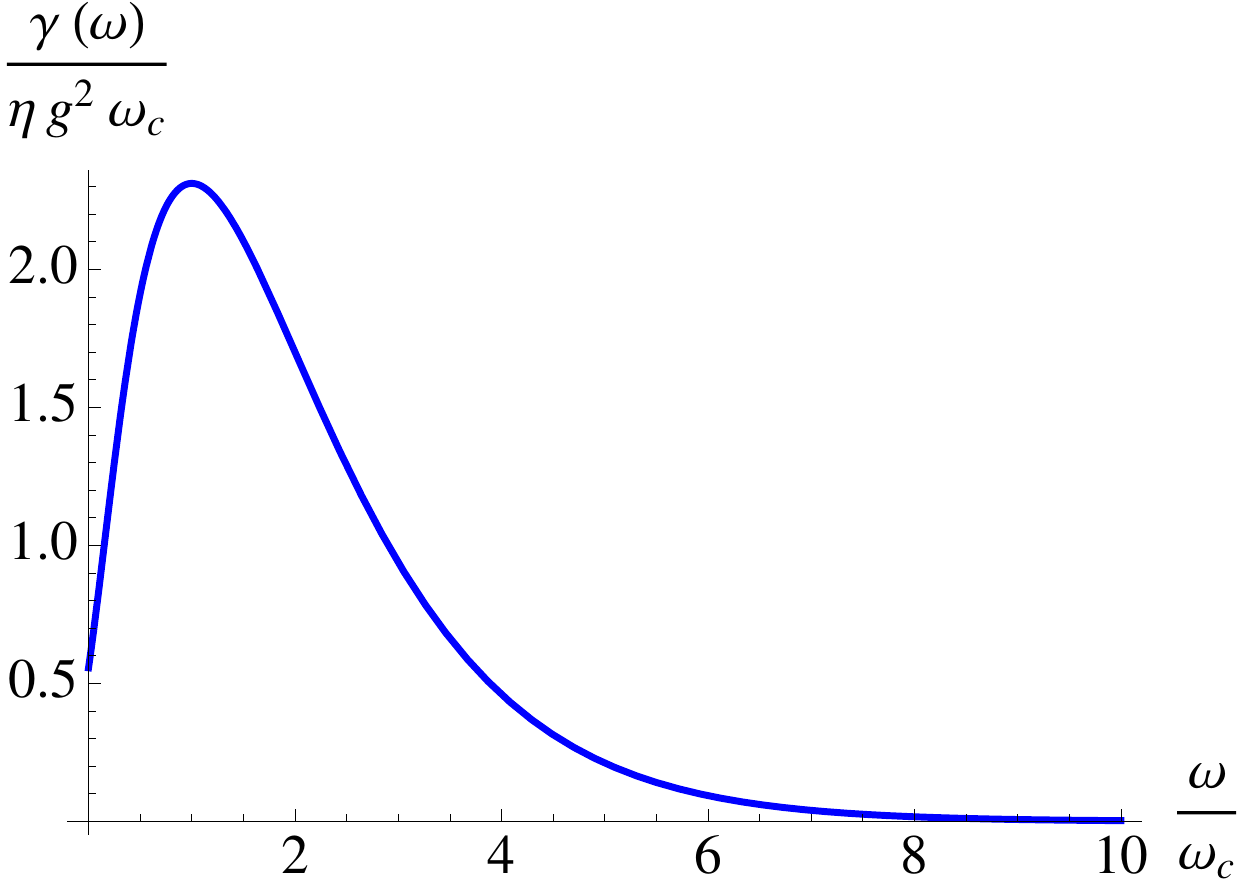} 
   \caption{Ohmic spectral density [Eq.~\eqref{eq:gamma}] with $1/\upbeta  = 2.23 \textrm{GHz}, \omega_c = 8 \pi \textrm{GHz}$.}
   \label{fig:gamma}
\end{figure}

The first summands in Eq.~\eqref{eq:32} (proportional to $\Gamma/[2\lambda^2(s)]$) are purely due to the evolving instantaneous energy, and are a factor of $t_f$ smaller than the remaining terms.  Therefore, for sufficiently large $t_f$ (i.e., when the adiabatic condition~\eqref{eq:36} is well satisfied), these terms can be neglected.   
It is simpler to analytically solve the dynamical equations \eqref{eq:32} in this limit, since it decouples the diagonal and off-diagonal elements. Using the KMS condition \eqref{eq:KMS} again to relate $\mathcal{F}_+(s)$ and $\mathcal{F}_-(s)$ we can rewrite these equations in the adiabatic limit as:
\bes
\label{eq:33}
\begin{align}
\frac{d}{ds} \rho_{--}(s) & = \mathcal{F}_+(s) \left[ 1 - \left( 1 + e^{-\upbeta \Delta(s)} \right) \rho_{--}(s) \right]  \ ,  \label{eqt:Edecoherence} \\
\frac{d}{ds} \rho_{+-}(s) & =  - \left[ i \Omega(s) +\Sigma(s) \right] \rho_{+ -}(s)  \ .  
\label{eqt:Edephasing}
\end{align}
\ees
These equations have solutions given by:
\bes
\begin{align}
\rho_{--}(s) & = \exp \left[- \int_0^s ds' \left( 1 + e^{-\upbeta \Delta(s')} \right) \mathcal{F}_+(s') \right] \times \notag \\
& \left( \rho_{--}(0) + \int_0^s ds' \mathcal{F}_+(s') \times \right. \notag \\
& \left. \exp \left[ \int_0^{s'} ds'' \left( 1 + e^{-\upbeta \Delta(s'')} \right) \mathcal{F}_+(s'') \right] \right) \ , \\
\rho_{+-}(s) & = \exp \left[ - \int_0^s ds' \left( i \Omega(s') + \Sigma(s') \right) \right] \rho_{+-}(0) \ ,\\
\rho_{++}(s) & = 1 - \rho_{--}(s) \ , \\
\rho_{-+}(s) & = \rho_{+-}^{\ast}(s) \ .
\end{align}
\ees
As is already clear from Eq.~\eqref{eqt:Edecoherence}, any deviation in the population of the instantaneous ground state, i.e., $\rho_{--}$, in the adiabatic limit is purely due to a non-zero $\mathcal{F}_+(s)$, which in turn requires a non-zero $\gamma(\Delta(s))$, i.e., a ``resonant" thermal excitation. This means that the rate of population loss is expected to be non-monotonic in the instantaneous energy gap $\Delta(s)$, i.e., when the goal is to prevent population loss from the ground state, \emph{increasing the gap can do damage before it starts to help}. The details of this depend on the decay rate $\gamma(\omega)$ via the bath correlation function $\mathcal{B}(t)$, as is clear from Eq.~\eqref{eq:gamma_ab} (see also Fig.~\ref{fig:gamma}). Also noteworthy is that the Lamb shift $S(\Delta(s))$ affects only the off-diagonal elements, through $\Omega(s)$.

Equation \eqref{eqt:Edephasing} highlights that, even in the time-dependent case, the decoherence occurs in the instantaneous energy eigenbasis.  In the adiabatic limit in which Eq.~\eqref{eq:33} is derived, the dynamics of the phase coherence between energy eigenstates completely decouples from that of the energy state populations, such that its exponential decay [at a rate determined by $\Sigma(s)$, Eq.~\eqref{eqt:Sigma}] does not affect the evolution of the energy state populations.  If $\rho_{+-}$ is initially zero, e.g., if the system is initialized purely in the ground state, then no coherence with the excited state is ever generated.   Therefore, we find that in the WCL, even in the presence of fast dephasing, an adiabatic computation is possible.  Similarly, Eq.~\eqref{eqt:Edecoherence} highlights that the thermal excitation/relaxation processes occur in the instantaneous energy eigenbasis since the dynamics is entirely determined by the population in the energy eigenstates

\begin{figure*}[t]
\begin{minipage}{0.48\textwidth}
   \subfigure[\ closed system, $t_f \omega_x = 10 \sqrt{2}$]{ \includegraphics[width=.9\columnwidth]{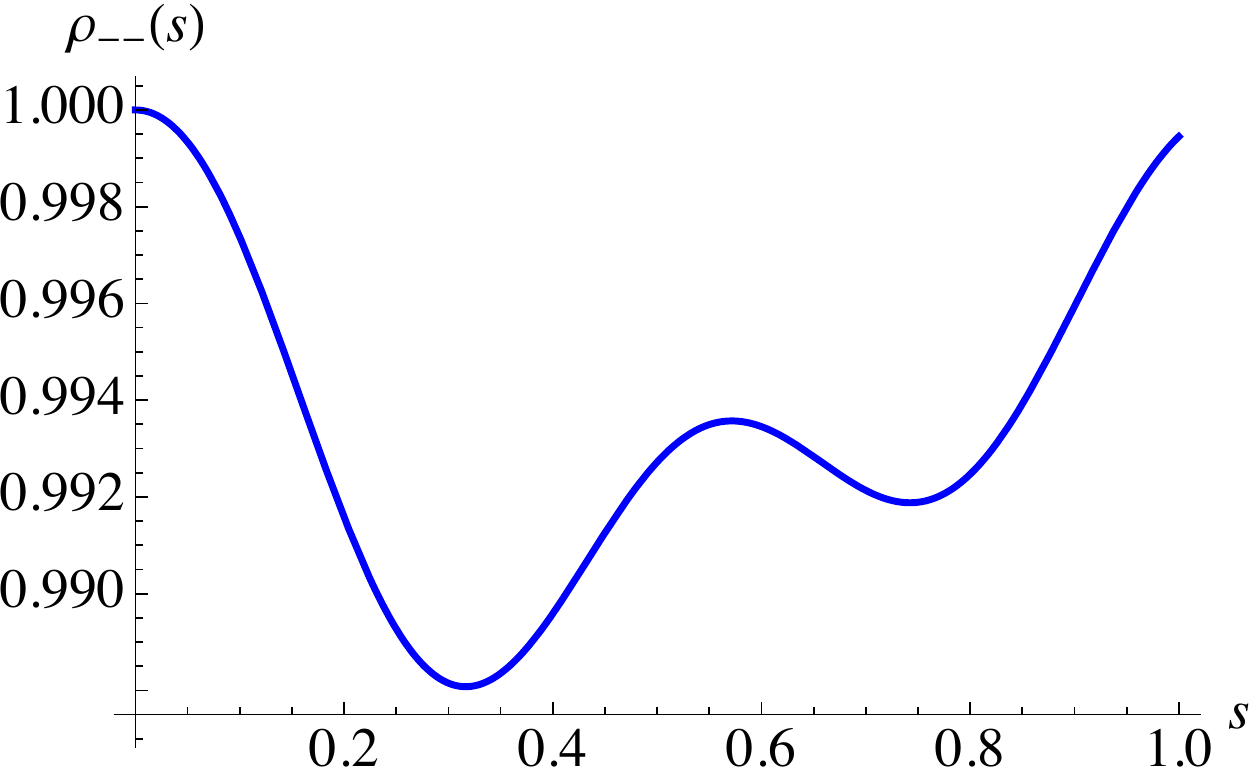}\label{fig:1Qubit_tf=14_closed}} 
   \subfigure [\ open system, $t_f \omega_x = 10 \sqrt{2}$]{ \includegraphics[width=.9\columnwidth]{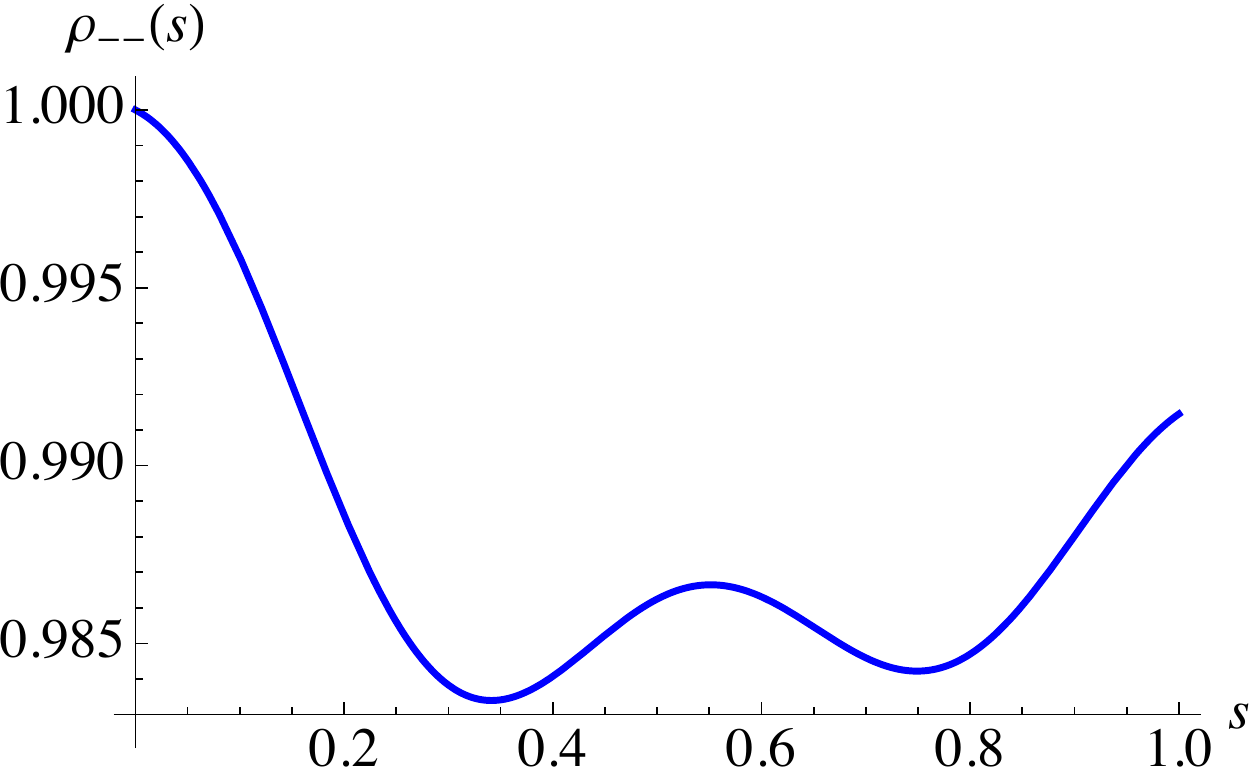} \label{fig:1Qubit_tf=14}}
\end{minipage}
\begin{minipage}[c][][t]{0.48\textwidth}
   \subfigure [\ open system, $t_f \omega_x = 5 \times 10^3$]{ \includegraphics[width=.9\columnwidth]{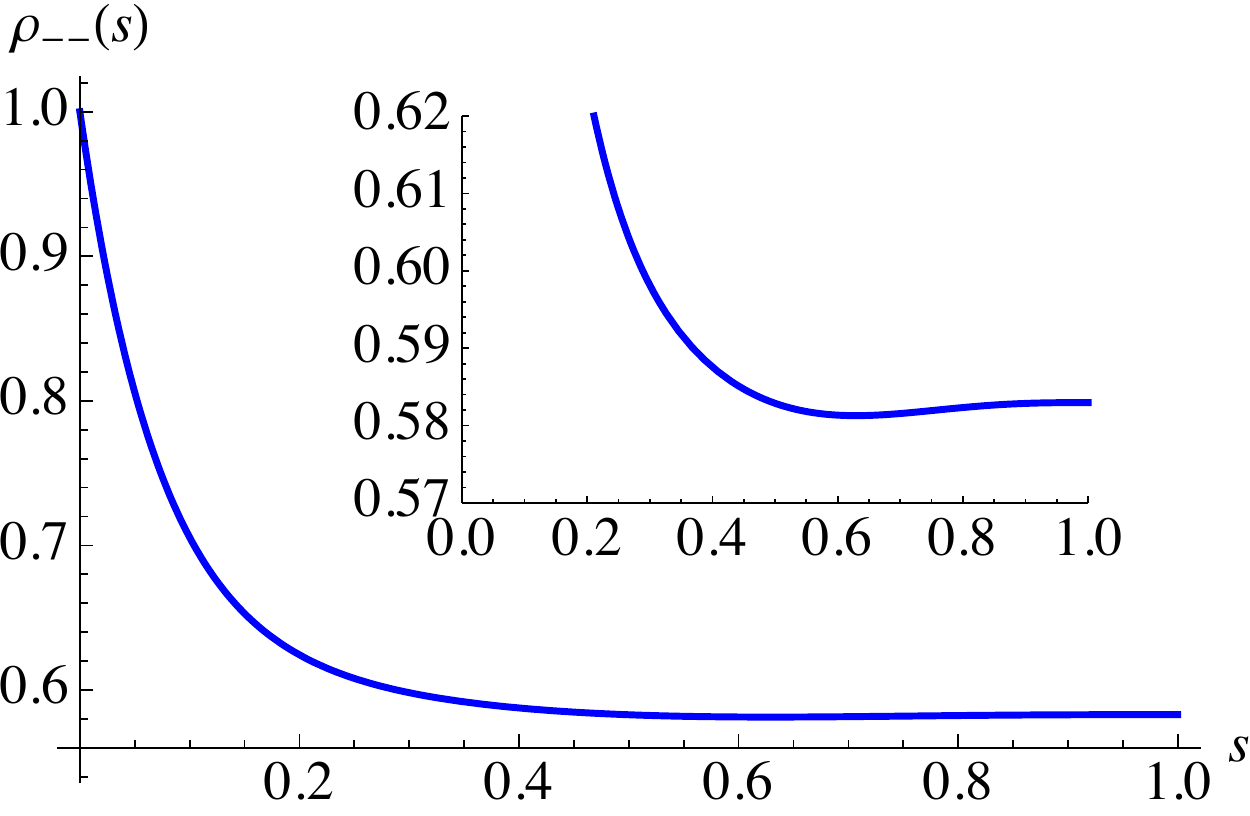} \label{fig:1Qubit_tf=5}}
   \subfigure [\ open system, $t_f \omega_x = 5 \times 10^4$]{ \includegraphics[width=.9\columnwidth]{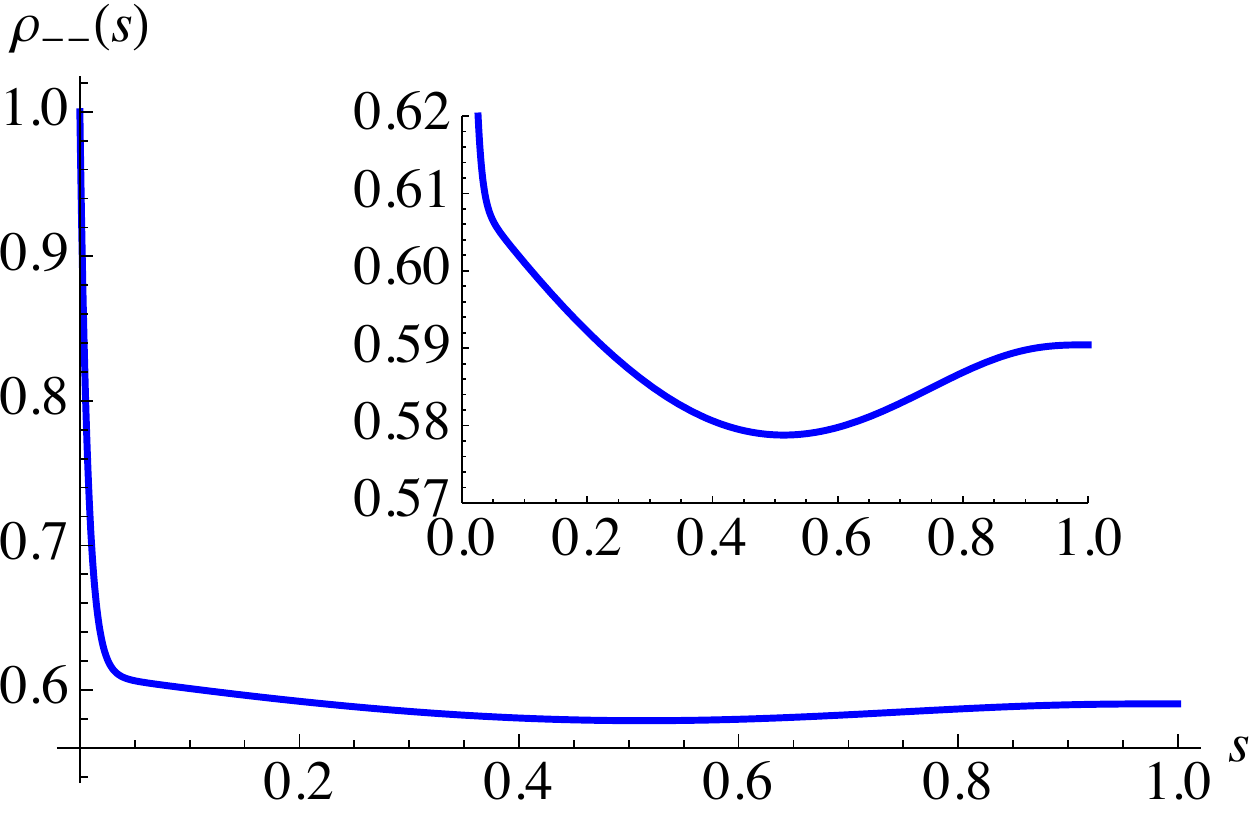}\label{fig:1Qubit_tf=50}}
\end{minipage}
\caption{Time-dependence of the ground state population in the WCL for a qubit evolving adiabatically subject to the Hamiltonian \eqref{eq:H_S-lin(t)} with $\rho(0) = \ket{\eps_-(0)}\bra{\eps_-(0)}$.  In (a) we depict the case of closed system evolution with $t_f \omega_x = 10 \sqrt{2}$, where the oscillation is entirely due to unitary non-adiabatic transitions. In (b)-(d) the system (with $t_f \omega_x = 10 \sqrt{2}$, $t_f \omega_x = 5 \times 10^3$, $t_f \omega_x = 5 \times 10^4$ respectively) is coupled to an Ohmic bath with $\gamma(\omega)$ as in Eq.~\eqref{eq:gamma}, and the instantaneous ground state population becomes gradually more damped as $t_f$ increases [the insets in (c) and (d) zoom in on the vertical axis]. The instantaneous ground state population initially decreases rapidly due to thermal excitation and then recovers somewhat due to thermal relaxation. In (b) the oscillation is a damped version of what is seen in (a). In (c) the oscillation is  completely damped by the bath, and a small population recovery is seen towards the end of the evolution. This recovery grows at even larger $t_f$ as seen in (d), indicating that this is due to thermal relaxation (this is shown in more detail in Fig.~\ref{fig:1Qubit_tf}). Parameters are chosen to satisfy the various inequalities: $\omega_x = \omega_z = 1$ GHz so $\Gamma =1$ and $\lambda_{\min} = 1/\sqrt{2}$ and Eq.~\eqref{eq:36} is satisfied since ${\Gamma}/{2\lambda_{\min}^3}=\sqrt{2} \ll t_f \omega_x$. For the Ohmic spectral density we chose ${\eta g^2} = 10^{-4}$,
$1/\upbeta  = 2.23 \textrm{GHz} < \omega_c = 8 \pi \textrm{GHz}$ (as in Fig.~\ref{fig:gamma}) so that Eq.~\eqref{eq:43} applies and is satisfied since $\left(\frac{\upbeta}{2\pi}\right)^2 \frac{\omega_z \omega_x}{2\lambda_{\min}} = \left(\frac{1}{2\pi 2.23}\right)^2 \frac{1}{\sqrt{2}} <1 \ll t_f \omega_x$.}
   \label{fig:1Qubit_Evolution}
\end{figure*}

\subsubsection{Solution without the adiabatic limit}
Let us now return to the full dynamical equations \eqref{eq:32} without taking the adiabatic limit.
Since $\mathcal{F}$, $\Omega$ and $\Sigma$ depend on $t_f$, there is a non-trivial dependence on $t_f$ for the final ground state population. To explore this dependence we solve the  dynamical equations \eqref{eq:32} numerically.  We
assume the bath is in a thermal state with an Ohmic spectral density, i.e., 
\beq
\gamma(\omega) = 2 \pi {\eta g^2} \frac{\omega e^{-|\omega|/\omega_c}}{1 - e^{-\upbeta \omega}}\ ,
\label{eq:gamma}
\eeq
where $\omega_c$ is a high-frequency cutoff and $\eta$ is a positive constant with dimensions of time squared arising in the specification of the Ohmic spectral function. This spectral density is depicted in Fig.~\ref{fig:gamma} for typical parameters used in our numerical calculations.
When the cutoff is the largest energy scale, specifically when $\omega_c \gg 1/\upbeta$, the bath correlation time [Eq.~\eqref{eq:tauB}] can be shown to be $\tau_B = \upbeta/(2\pi)$ \cite{ABLZ:12-SI}. Thus the validity condition \eqref{eq:8c} becomes, in addition to Eq.~\eqref{eq:36}:
\bes
\label{eq:43}
\begin{align}
& t_f \gg \left(\frac{\upbeta}{2\pi}\right)^2 \frac{\omega_z}{2\lambda_{\min}}  \quad \textrm{or, equivalently}  \\  
& t_f \sqrt{\omega_x \omega_z} \gg \frac{1}{2} \left(\frac{\upbeta}{2\pi}\right)^2 \sqrt{\Gamma \omega_x^4+\Gamma^{-1} \omega_z^4}\ .
\end{align}
\ees
\begin{figure}[t] 
   \centering
   \includegraphics[width=0.9\columnwidth]{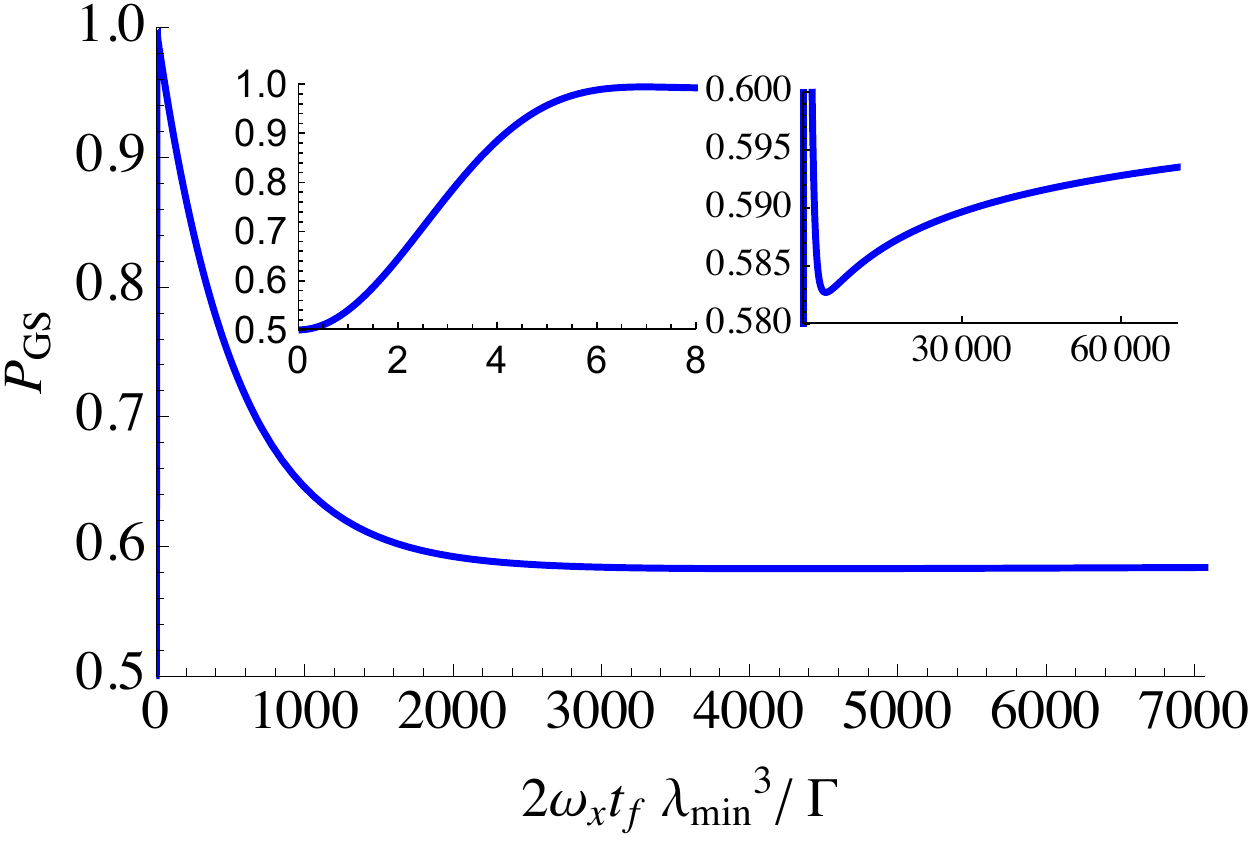} 
\caption{Final WCL ground state population ($P_{\textrm{GS}} =\rho_{--}(t_f)$) as a function of evolution time [in dimensionless units corresponding to the adiabatic condition~\eqref{eq:36}] for the model specified in Fig.~\ref{fig:1Qubit_Evolution} with the same initial condition.  The left inset zooms in on the short total-time evolution, the right inset shows the long total-time evolution. The maximum in the ground state probability (left inset) is the optimal evolution time and is seen to occur at $2t_f \omega_x \lambda_{\min}^3/{\Gamma}\approx 7$, i.e., increasing $t_f$ much above the heuristic adiabatic condition does not help. The reason is that this maximum is a balance between maximizing adiabaticity while minimizing thermal excitations. The right inset shows that for very long evolution times thermal relaxation repopulates the ground state, which eventually settles on its Gibbs distribution value. The condition~\eqref{eq:43} [expressed as $\left(\frac{\upbeta}{2\pi}\right)^2 \frac{\omega_z}{2t_f\lambda_{\min}}=1$] is satisfied already at $2t_f \omega_x \lambda_{\min}^3/{\Gamma}\approx 10^{-3}$.}
   \label{fig:1Qubit_tf}
\end{figure}

Examples of evolutions satisfying Eq.~\eqref{eq:43} are shown in Fig.~\ref{fig:1Qubit_Evolution} for both the closed system case and the open system case for increasing values of $t_f$. It shows that oscillations of the instantaneous ground state population due to non-adiabatic unitary dynamics are damped out as the total evolution increases, and that some ground state population is recovered due to the thermal relaxation for large enough $t_f$. The dependence of the final population on $t_f$ is shown in Fig.~\ref{fig:1Qubit_tf}, where we explicitly see the effect of the various timescales in our problem.  For very short evolution times (where the adiabatic condition is not satisfied, i.e., $2t_f \omega_x \lambda_{\min}^3/\Gamma < 1$, the evolution is highly non-adiabatic, and the final ground state probability is close to $1/2$ (see the left inset of Fig.~\ref{fig:1Qubit_tf}). However, note that in this regime we may not entirely trust the ME to be a reliable approximation for the dynamics since the condition $h/t_f \ll \tau_B^{-2}$ [Eq.~\eqref{eq:8c}] requiring that the system evolves much more slowly than the time-scale of the bath is not necessarily satisfied.  Furthermore, since the evolution is so short that $\mathcal{F}_+(s) \leq t_f \gamma(\Delta(0)) = t_f \gamma(\omega_x) \ll 1$ [recall Eq.~\eqref{eq:32a}], thermal effects are small because they have insufficient time to act.  

As we increase $t_f$, the evolution becomes more and more adiabatic and the ground state probability peaks close to $1$ around $2t_f \omega_x \lambda_{\min}^3/{\Gamma}=1$, the adiabatic condition~\eqref{eq:36} (see first inset of Fig. \ref{fig:1Qubit_tf}).  However, as we continue to increase $t_f$, rather than observing that the system remains in its ground state, thermal excitations increase; this removes significant population from the ground state, which actually drops below its thermal equilibrium probability distribution.  An example of this is shown in Fig.~\ref{fig:1Qubit_tf=5} where the ground state population decreases over the first half of the evolution.  As we continue to increase $t_f$, thermal relaxation allows the system to relax to its thermal equilibrium probability distribution.  For an example of the effect of thermal relaxation, see the increase in ground state probability during the second half of the evolution in Fig.~\ref{fig:1Qubit_tf=50}.  The increase in the population as we increase $t_f$ is shown in the second inset of Fig.~\ref{fig:1Qubit_tf}. 

\begin{figure}[t] 
   \centering
   \includegraphics[width=0.9\columnwidth]{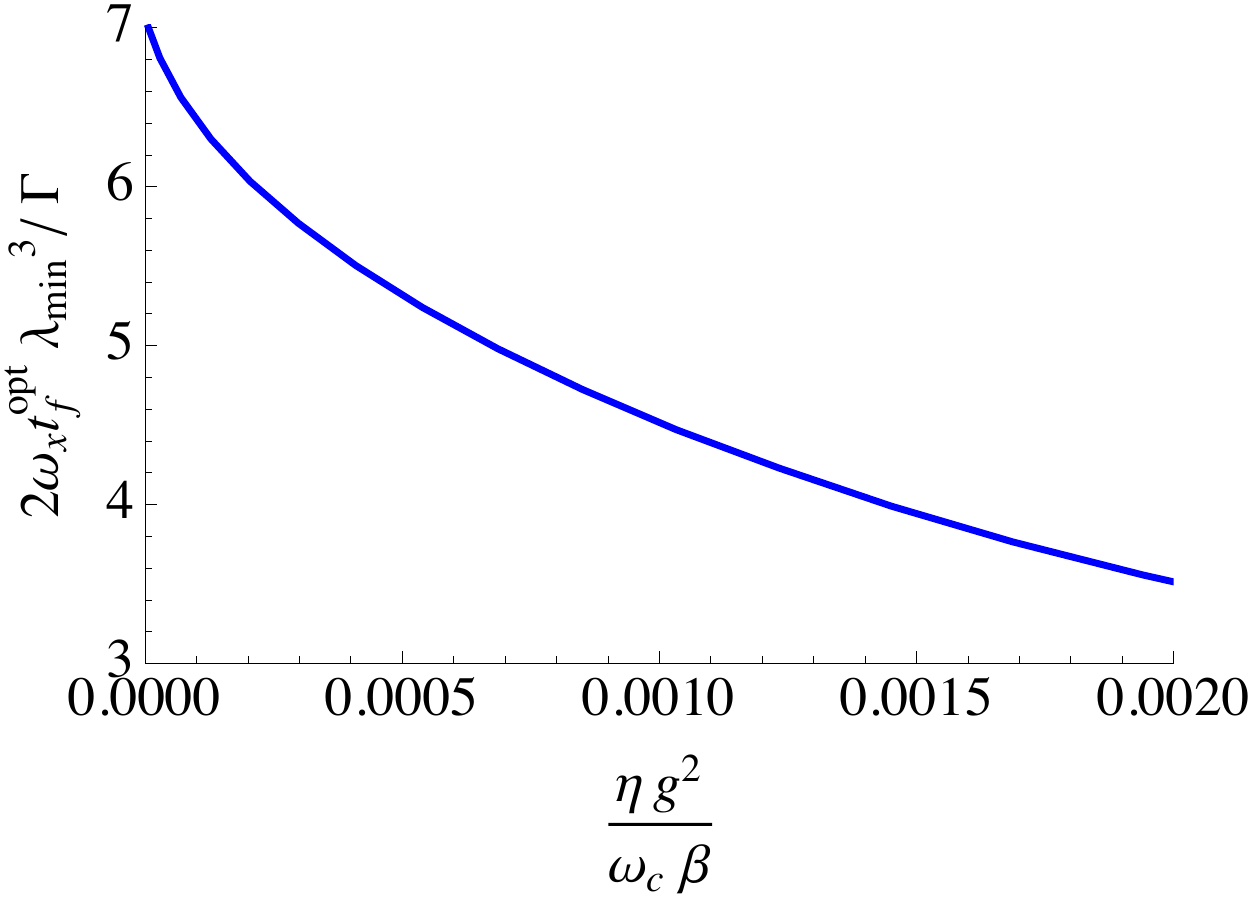} 
\caption{Behavior of the the ``adiabatic parameter'' defined in Eq.~\eqref{eq:36} at the optimal evolution time $t_f^{\textrm{opt}}$ (that maximizes the ground state probability) on the system-bath coupling.  As the coupling strength increases, the optimal evolution time becomes smaller, and the evolution becomes less adiabatic.  Simulation parameters: $\omega_x = 1 \textrm{GHz}$, $\Gamma = 1$, $1/\upbeta  = 2.23 \textrm{GHz}$, $\omega_c = 8 \pi \textrm{GHz}$. Note that the adiabatic parameter never dips below $1$ (which would violate the heuristic adiabatic condition): $t_f^{\textrm{opt}}$ no longer exists past the last data point shown because beyond that point the highest $P_{\textrm{GS}}$ achievable is the thermal one, which cannot be exceeded no matter how large $t_f$ becomes. The smallest system-bath coupling plotted is $\frac{\eta g^2}{\omega_c \beta} = 5.67 \times 10^{-6}$.}
   \label{1Qubit_optimal}
\end{figure}

Thus thermal excitations can have a significant detrimental impact on the ground state population, and hence the success probability of an adiabatic quantum computation.  As can be seen in Fig.~\ref{fig:1Qubit_tf}, there is an optimum value for $t_f$ that maximizes the ground state probability.  This value balances the adiabaticity of the evolution against the time allowed for thermal processes to occur.  We illustrate the dependence of this optimal time on the bath strength in Fig.~\ref{1Qubit_optimal}, where we see that as the system-bath strength increases in value, such that thermal processes occur more rapidly, the optimal evolution time decreases.  This shows that in open system adiabatic quantum computation it can be advantageous to stop the computation early, before the ground state probability starts to dip due to thermal excitations, in agreement with earlier findings \cite{Steffen:2003ys,PhysRevA.71.012331,ABLZ:12-SI,PhysRevLett.95.250503,crosson2014different}.

\begin{figure*}[t]
\begin{center}
\subfigure[\ $\omega_x t_f = 10^2$]{\includegraphics[width=0.32\textwidth]{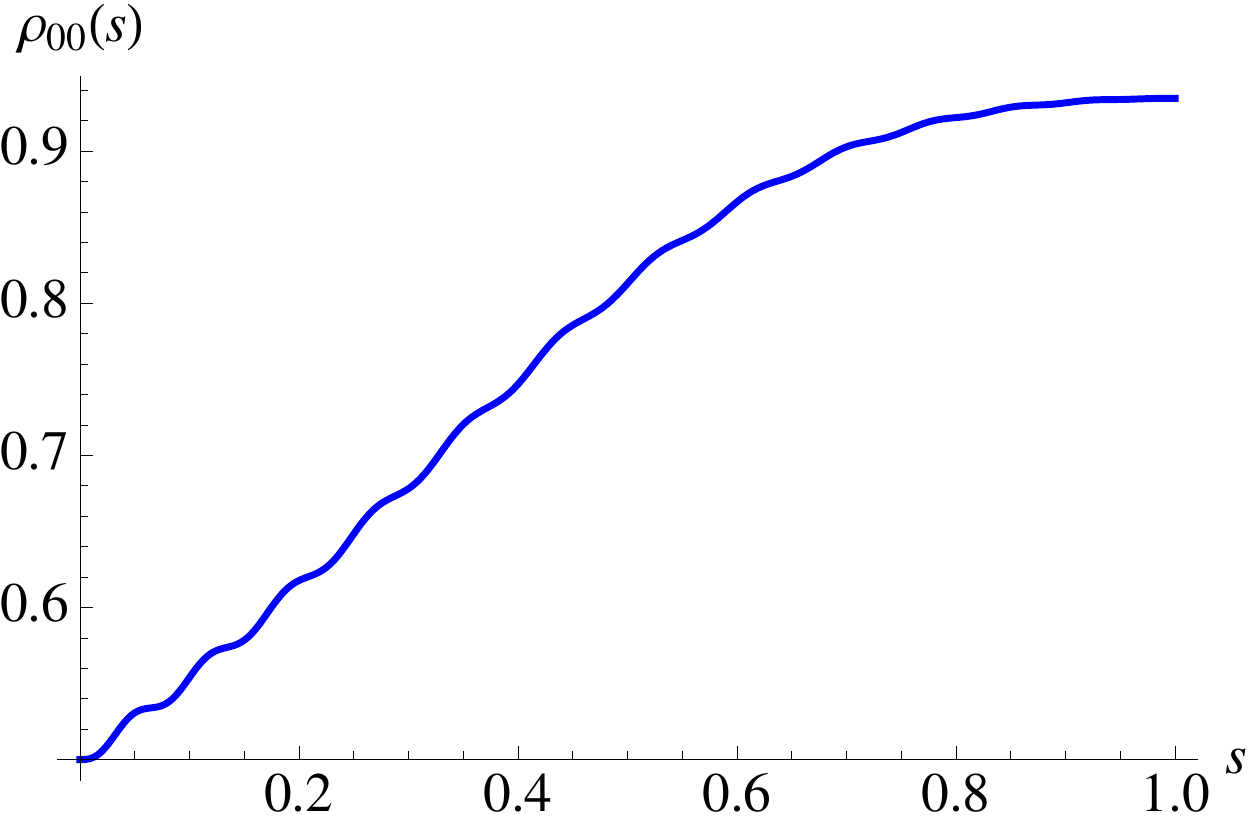} \label{fig:1Qubit_SCL_tf=100}} 
\subfigure[\ $\omega_x t_f = 10^3 $]{\includegraphics[width=0.32\textwidth]{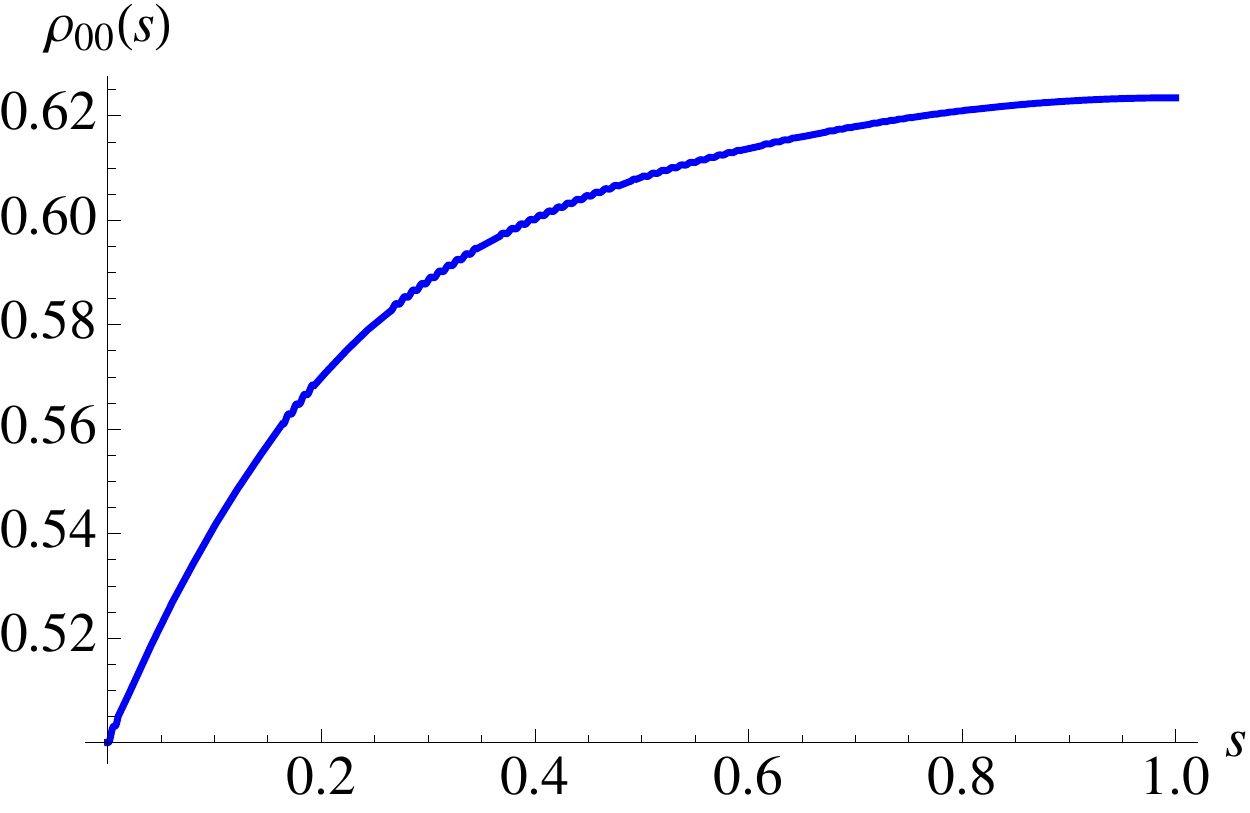} \label{fig:1Qubit_SCL_tf=1000}} 
\subfigure[\ $\omega_x t_f = 10^4$]{\includegraphics[width=0.32\textwidth]{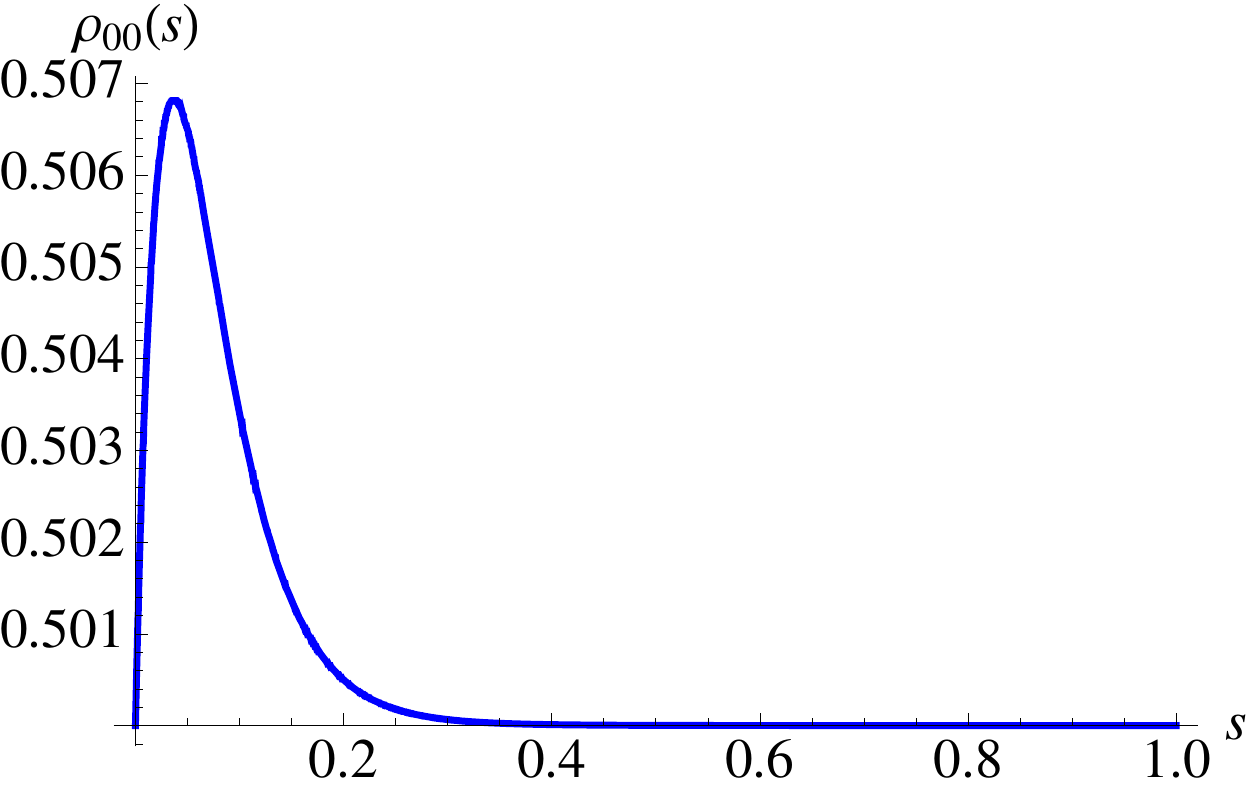} \label{fig:1Qubit_SCL_tf=10000}}
\end{center}
\caption{Time-dependence of the ground state population in the SCL for a qubit evolving adiabatically subject to the Hamiltonian \eqref{eq:H_S-lin(t)} and coupled to an Ohmic bath with $\gamma(\omega)$ as in Eq.~\eqref{eq:gamma}, yielding $\gamma(0) = 2 \pi{\eta g^2}/\beta$.  The initial state is $\rho(0) = \ket{\eps_-(0)}\bra{\eps_-(0)}=\ketbra{+}{+}$.  Depicted is the numerical solution of Eq.~\eqref{eq:44}. In (a) $\omega_x t_f = 10^2$ such that $\Gamma/ ( \omega_x \lambda_{\min}^3) \ll t_f \ll T_2^{(c)}$ so that the evolution is adiabatic and unaffected by decoherence, and the ground state population increases to $1$ in the computational basis. In (b) $\omega_x t_f = 10^3 $ such that $\Gamma/ ( \omega_x \lambda_{\min}^3) \ll t_f \lesssim  T_2^{(c)}$, resulting in a significant loss of ground state population. In (c) $\omega_x t_f = 10^4$ such that $t_f \gg  T_2^{(c)}$ and the ground state population rapidly settles on $1/2$. The temperature and spectral density parameters are as in Fig.~\ref{fig:1Qubit_Evolution}.}
   \label{fig:1QubitSCL_Evolution}
\end{figure*}
%

\subsection{SCL}
\label{sec:SCL2}
We can perform a similar analysis in the case of the SCL, Eq.~\eqref{eqt:SCL}.  We find that the dynamical equations are given by: 
\bes
\label{eq:44}
\begin{align}
\frac{d}{ds} \rho_{0 0} & = -\frac{d}{ds} \rho_{1 1} = {-t_f \omega_x \frac{i}{2}(1-s)\left(  \rho_{0 1} -   \rho_{1 0 } \right)}  \ , \\
\frac{d}{ds} \rho_{0 1} & = \frac{d}{ds} \rho_{10}^\ast   =  - \frac{t_f}{T_2^{(c)}}  \rho_{0 1}+ \notag \\
&\quad t_f \omega_x {\frac{i}{2 } \left[  (1-s) \left( \rho_{1 1} - \rho_{0 0} \right) + 2 s \Gamma \rho_{0 1} \right]} . 
\end{align}
\ees
\begin{figure}[t] 
   \centering
   \includegraphics[width=0.95\columnwidth]{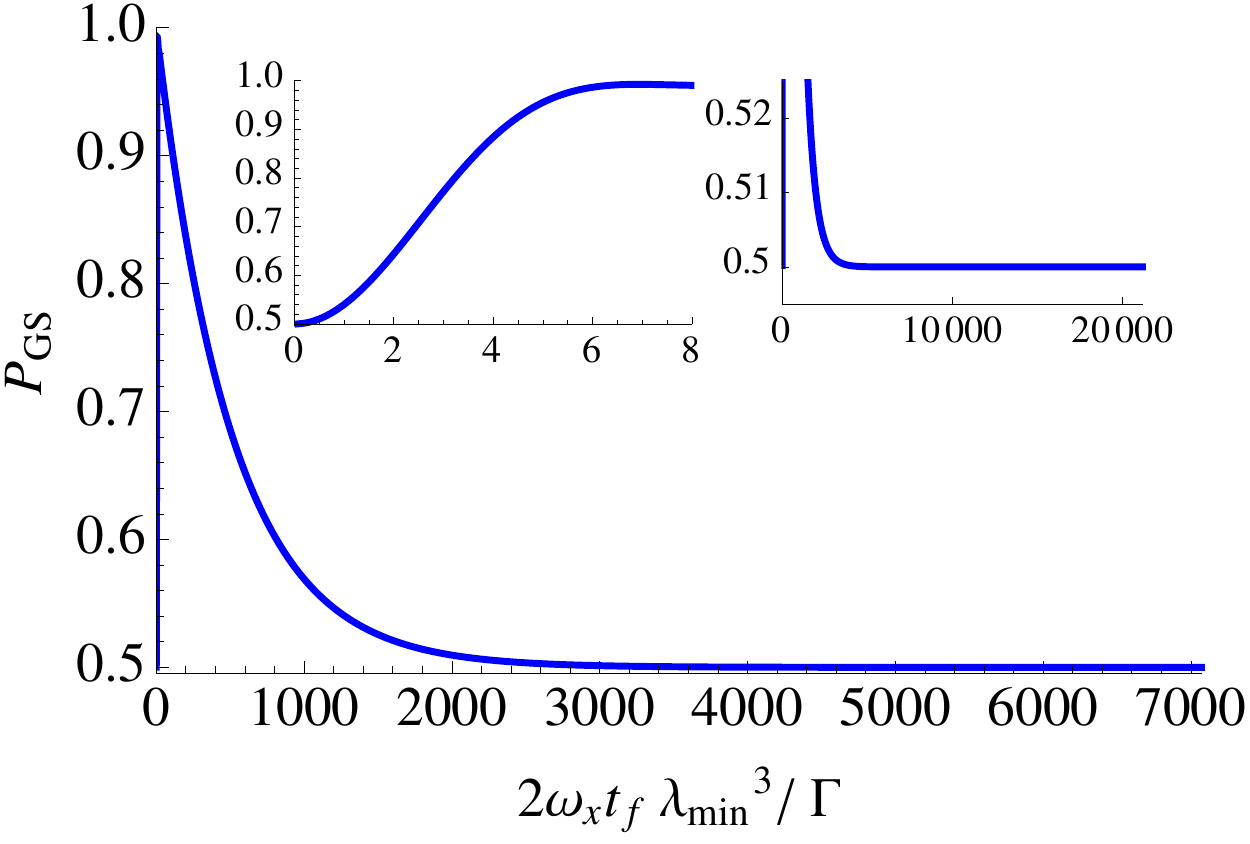} 
   \caption{Final SCL ground state population [$P_{\textrm{GS}} =\rho_{00}(t_f)$]  as a function of evolution time. The left inset zooms in on the short total-time evolution, the right inset shows the long total-time evolution.  For these long total-time evolutions, the system state rapidly becomes maximally mixed. The parameters are as in Fig.~\ref{fig:1Qubit_Evolution}.}
   \label{fig:1QubitSCL_tf}
\end{figure}
We plot their numerical solution in Figs.~\ref{fig:1QubitSCL_Evolution} and \ref{fig:1QubitSCL_tf}. With the initial condition being a fully populated  ground state, we observe behavior that depends strongly on the various timescales.  For $t_f \ll T_2^{(c)}=[2 \gamma(0)]^{-1}$ but $t_f \gg \Gamma/ ( \omega_x \lambda_{\min}^3)$ [the adiabatic condition~\eqref{eq:36}], the decoherence in the computational basis does not have enough time to disrupt the adiabatic evolution, and the system can perform an adiabatic computation. This is illustrated in Fig.~\ref{fig:1Qubit_SCL_tf=100}, where the population in the ground state at the end of the evolution is almost $1$.  But when $t_f  \gtrsim T_2^{(c)}$ [Fig.~\ref{fig:1Qubit_SCL_tf=1000}], even if the adiabatic condition is well satisfied, the off-diagonal entries of the density matrix start to decay exponentially and we begin to see a significant loss in the ground state population at the end of the evolution.  For $t_f \gg T_2^{(c)}$, the off-diagonal terms rapidly decay to zero, and the evolution results in a state that resembles the maximally mixed state [Fig.~\ref{fig:1Qubit_SCL_tf=10000}]. In this case the evolution is very well approximated by the results of the time-independent case in Eq.~\eqref{eqt:SCL_sol}.  The dependence on the final ground state probability is shown in Fig.~\ref{fig:1QubitSCL_tf}, and is seen to rapidly tend to the maximally mixed state value of $1/2$.

\begin{figure}[b] 
   \centering
   \includegraphics[width=0.95\columnwidth]{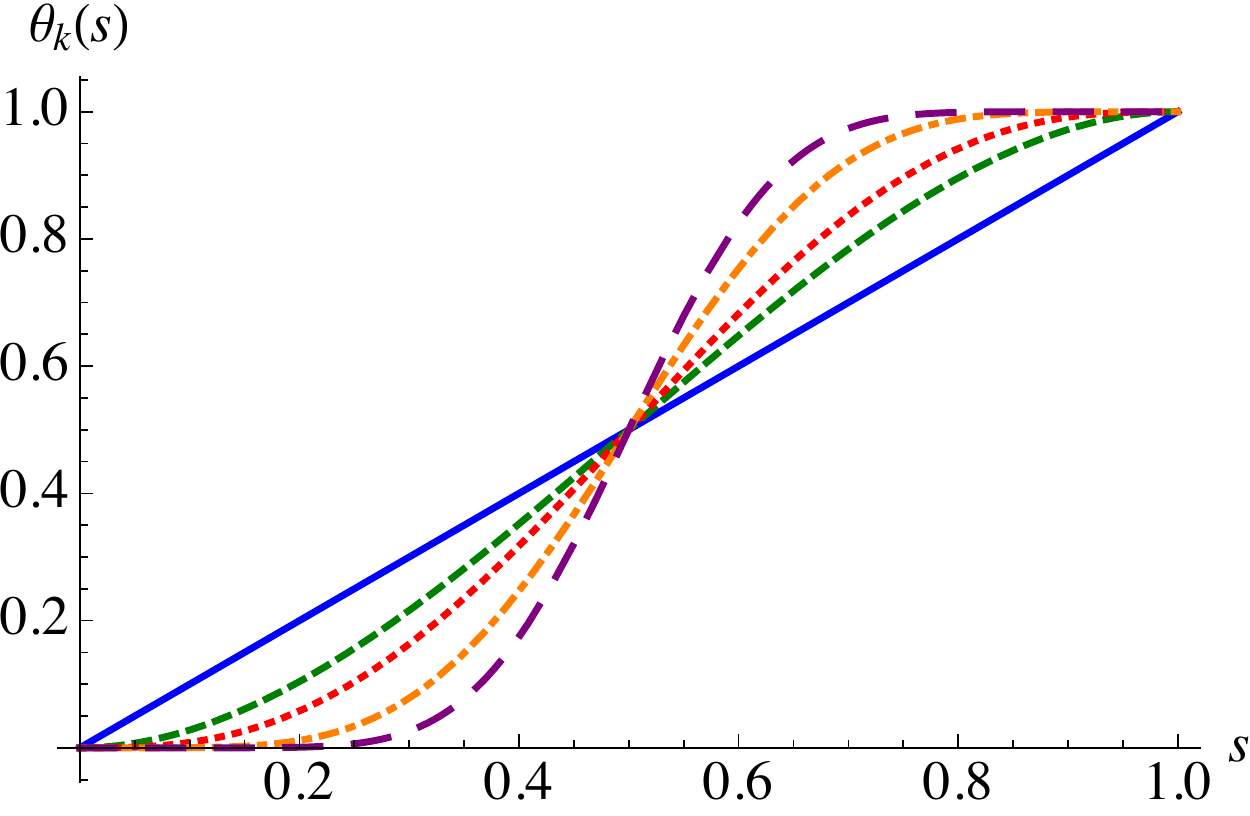} 
   \caption{The regularized incomplete beta functions [Eq.~\eqref{eq:beta}] for $k=0$ (blue-solid, linear), $1$ (green-dashed), $2$ (red-dotted), $5$ (orange-dot-dashed), $10$ (purple-long dashed, steepest rise).}
   \label{fig:Betafunctions}
\end{figure}
\begin{figure*}[t]
   \centering
\subfigure[\ $\eta g^2 = 0$]{\includegraphics[width=0.32\textwidth]{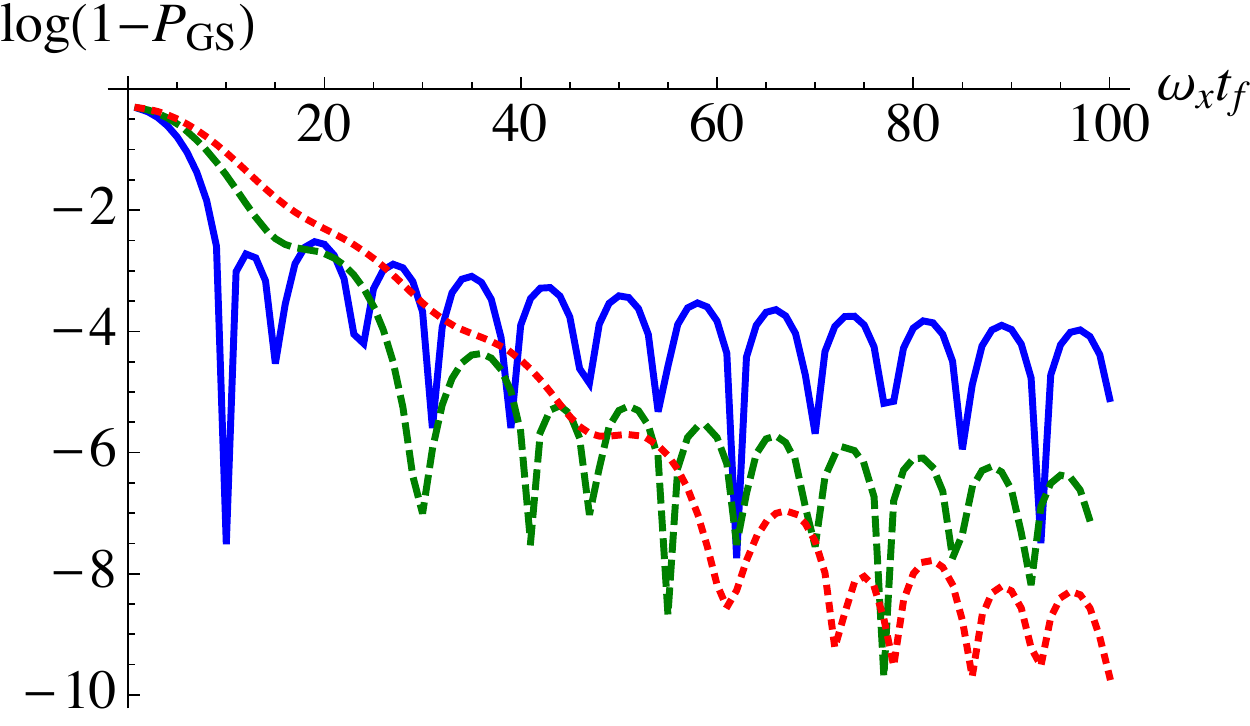} \label{fig:Beta_Closed}} 
\subfigure[\ $\eta g^2 = 10^{-8}$]{\includegraphics[width=0.32\textwidth]{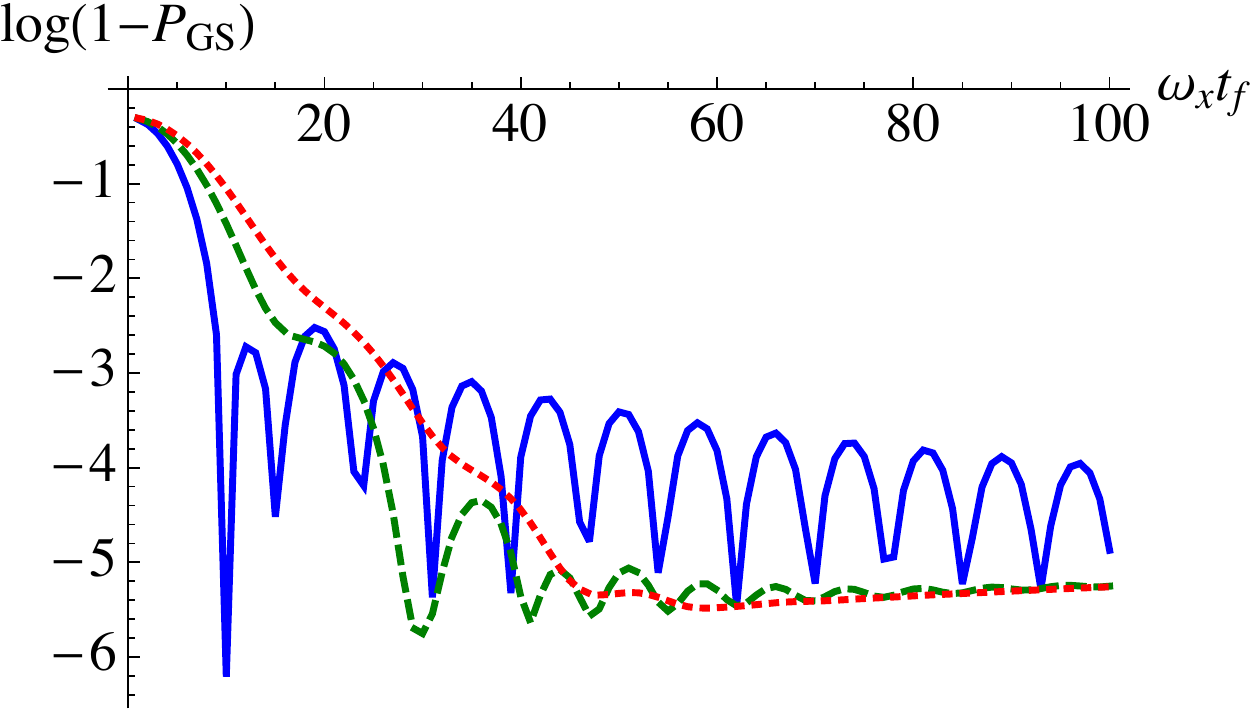} \label{fig:Beta_Open_g=10-4}} 
\subfigure[\ $\eta g^2 = 10^{-4}$]{\includegraphics[width=0.32\textwidth]{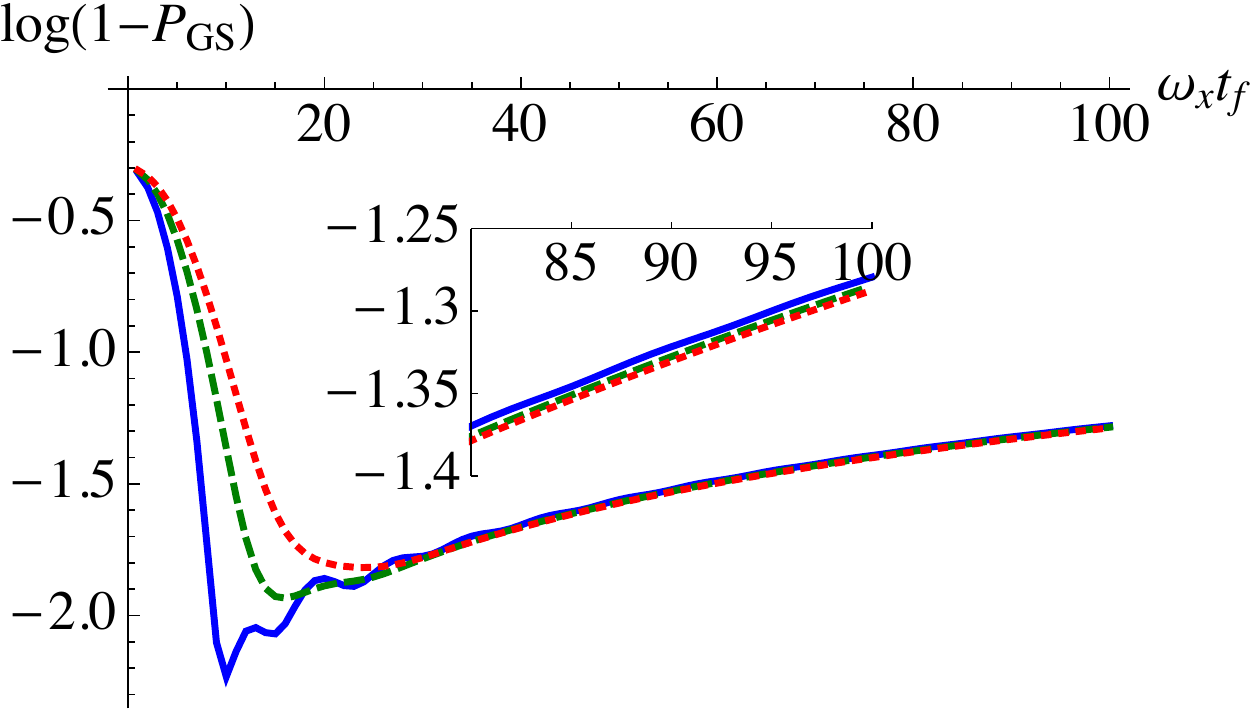} \label{fig:Beta_Open_g=10-2}}
\caption{Log base $10$ of the ground state error ($1$ minus the ground state probability) as a function of the total evolution time $t_f$ (in units of $\omega_x$) for different regularized incomplete beta function [Eq.~\eqref{eq:beta}] values of $k=0$ (blue, solid), $k=1$ (green, dashed), $k=2$ (red, dotted). The three cases shown correspond to (a) the closed system case, $\eta g^2 = 0$, (b) weak system-bath coupling, $\eta g^2 = 10^{-8}$, (c) somewhat stronger system-bath coupling, $\eta g^2 = 10^{-4}$. The efficacy of the boundary cancellation method decreases as the system-bath coupling strength grows.}
   \label{fig:BetaGS}
\end{figure*}

\subsection{WCL \textit{vs} SCL}
Concluding this study of the single qubit case, we observe some important differences between the WCL and the SCL. Most notably, as is apparent from comparing Figs.~\ref{fig:1Qubit_tf} and \ref{fig:1QubitSCL_tf}, while superficially the overall behavior of the ground state population appears qualitatively similar (especially for short $t_f$), the key difference is that in the WCL the final ground state population settles on a finite-temperature thermal equilibrium value, while in the SCL it approaches the infinite temperature (maximally mixed state) thermal equilibrium value of $1/2$. We shall see these conclusions reinforced below, when we discuss the multi-qubit case. The implication is that adiabatic quantum computation is possible in the WCL (since there is a non-vanishing probability for the system to end up in its ground state), while it is hopeless in the SCL where the final equilibrium value is $1/2^N$, where $N$ is the system size (number of qubits). The explanation for these conclusions is that in the case of the WCL the relevant timescales are the adiabatic and \emph{thermal relaxation} timescales.  The dephasing timescale does not matter since dephasing in the energy eigenbasis is not detrimental to the process of finding the final ground state.  
In contrast, in the case of the SCL the relevant timescales are the adiabatic and \emph{dephasing} timescales.  Since the dephasing is occurring in the computational basis, this destroys the coherence of the energy eigenstates.  

\section{Boundary Cancellation Method}
\label{sec:interp}
%
\begin{figure}[b] 
   \centering
   \includegraphics[width=0.95\columnwidth]{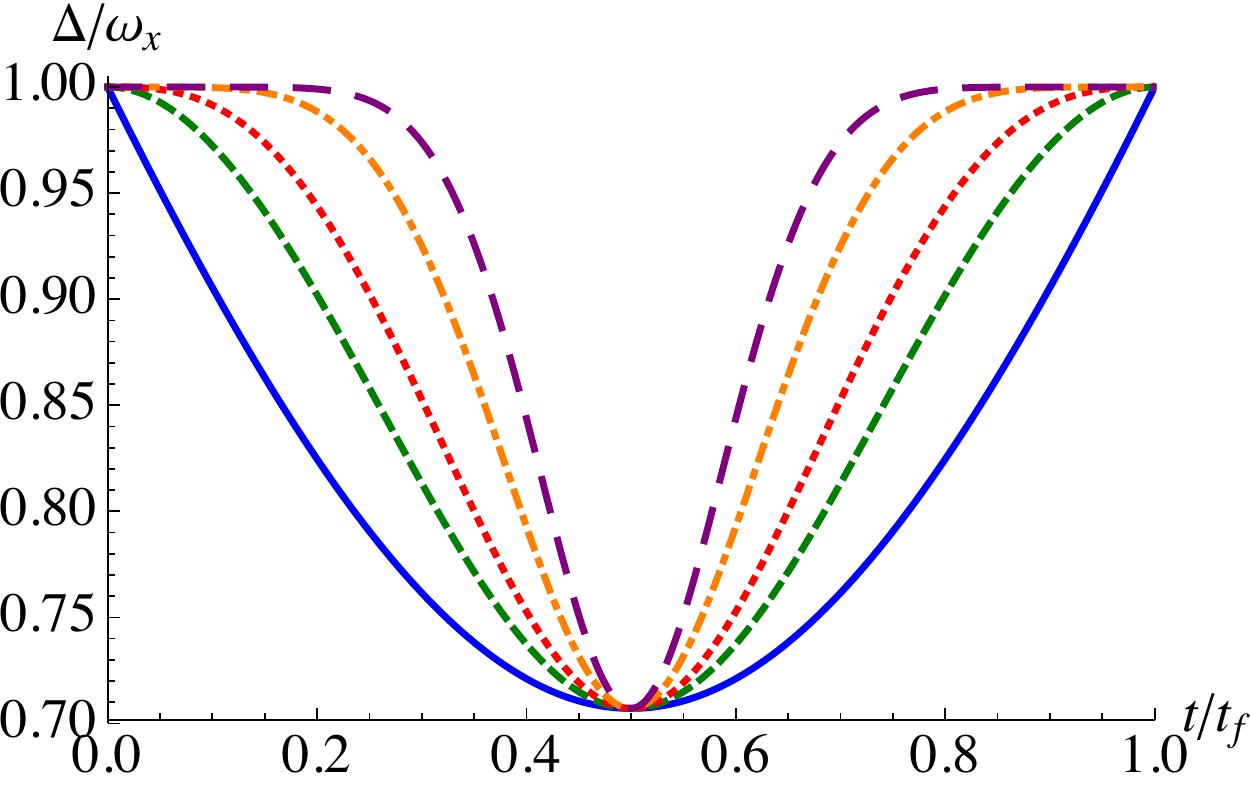} 
   \caption{The energy gap between the ground state and first excited state for the annealing schedules parametrized by $k=0,1,2,5,10$ (blue-solid, green-dashed, red-dotted, orange-dot-dashed, purple-long dashed).  The minimum gap remains unchanged, but the gap remains large for longer as $k$ increases.}
   \label{fig:BetaGap}
\end{figure}
\begin{figure}[b]
   \centering
\includegraphics[width=0.32\textwidth]{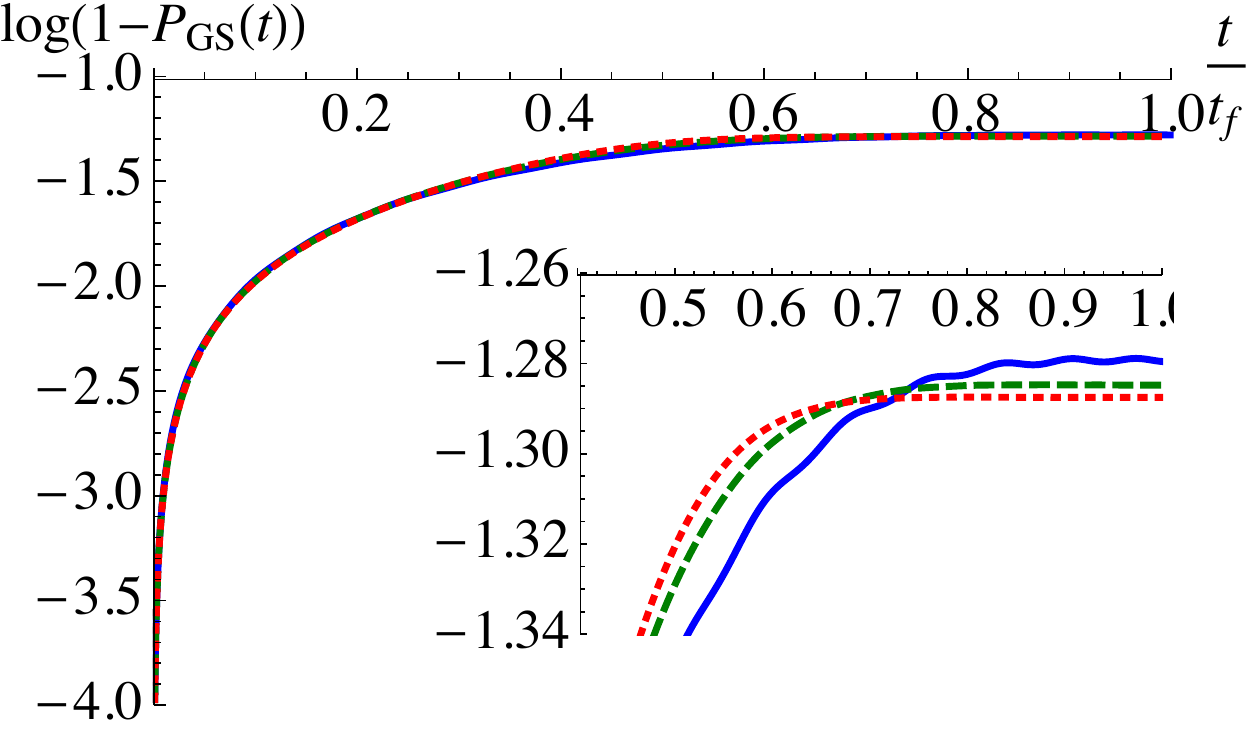}
\caption{Evolution of the instantaneous ground state error for different regularized incomplete beta function values of $k=0$ (blue-solid), $k=1$ (green-dashed), $k=2$ (red-dotted). We set $\eta g^2 = 10^{-4}$, and $\omega_x t_f = 100$.}
   \label{fig:BetaEvolutionGS}
\end{figure}
So far, we have only considered a linear interpolating function between the initial and final Hamiltonian. However, it is well known that optimization of the interpolation function can have important consequences. For example, it is only after such an optimization, based on a time-local adiabatic condition, that the quadratic speedup of Grover's algorithm was shown to be realizable in AQC \cite{Roland:2002ul}. Subsequently various studies have shown that the interpolating function can be optimized to improve the performance of the adiabatic algorithm in the closed system setting \cite{lidar:102106,PhysRevLett.103.080502,RPL:10,Wiebe:12}. Related open system results have also been reported \cite{PhysRevA.82.040304}. In particular, Ref.~\cite{lidar:102106} showed that in the closed system case, the deviation from the final ground state can be made arbitrarily small at a fixed $t_f$ (large enough to satisfy the adiabatic condition with respect to the minimum energy gap) by choosing a family of analytic interpolation functions with an increasing number $k$ of vanishing derivatives at the boundaries $t=0$ and $t=t_f$. In a nutshell, each such vanishing derivative cancels another boundary term in an integration-by-parts version of adiabatic perturbation theory. In Ref.~\cite{RPL:10} an explicit example of such an interpolation function was provided, again for the closed system setting, which we revisit here in the open system setting. Specifically, Ref.~\cite{RPL:10} proposed the interpolating function to be given by the regularized incomplete beta function $\theta_k(s)$:
\beq
\theta_k(s) = \frac{B_{s}(1+k,1+k)}{B_1(1+k,1+k)} \ ,
\label{eq:beta}
\eeq
where $B_s(a,b) = \int_0^s dy \ y^{a-1}(1-y)^{b-1}$ with $\Re(a), \Re(b) >0$ and $|s| \leq 1$.  A sample of the behavior of the family parametrized by $k$ (the number of vanishing derivatives at the boundaries) is shown in Fig.~\ref{fig:Betafunctions}. We again consider a single qubit in the WCL with the system Hamiltonian
\beq
H_S(t) = - \frac{1}{2}{\omega_x} \left[ 1 -\theta(s) \right] \sigma^x - \frac{1}{2} \theta(s) \omega_z \sigma^z \ ,
\label{eq:H_S-theta}
\eeq
coupled to an Ohmic bath, as in Sec.~\ref{sec:WCL2}. Figure~\ref{fig:BetaGS} displays the behavior of the ground state probability as we increase $k$ starting from zero (linear interpolation).  We observe that even in the open system case increasing $k$ leads to an improvement in the ground state probability in the adiabatic regime, though the improvement rapidly saturates as $k$ grows. Another attractive feature of increasing $k$, visible in Fig.~\ref{fig:BetaGS}, is that it suppresses the oscillations in the ground state probability.

It is interesting to check whether the improvement in the ground state probability is related to a change in the spectrum induced by the choice of interpolation function. As shown in Fig.~\ref{fig:BetaGap}, the closed-system minimum energy gap remains unchanged by changing $k$, but increasing $k$ results in the gap remaining large for a longer time. This, in turn, is correlated with a reduction in the amount of thermal excitations, as shown in Fig.~\ref{fig:BetaEvolutionGS}. 

Thus, the boundary cancellation method has a positive, albeit mild effect, even for open systems. We next turn our attention to the multi-qubit case.
\section{The multi-qubit case}
\label{sec:VII}
%
We now extend our discussion to the multi-qubit case, where the differences between the SCL and the WCL become most prominent. In particular, we shall address the key question posed in the introduction: how exactly does the single-qubit $T_2$ enter, and how detrimental is a short $T_2$ for successful AQC? We shall see that the answer depends dramatically on whether the SCL or WCL applies.
Namely, we shall demonstrate that in the SCL with independent baths, the decoherence time can scale with the system size and hence rapidly destroy the coherence of energy eigenstates.  This renders any computation in AQC effectively impossible in the SCL as the system size grows, without extensive error correction. On the other hand, we shall demonstrate that even in the multi-qubit case,  AQC remains possible in the WCL limit, with the dominant source of error being thermal relaxation to the finite-temperature thermal state, but with the single-qubit $T_2$ not playing an important role.

\subsection{SCL}
Comparing Figs.~\ref{fig:1Qubit_tf} and \ref{fig:1QubitSCL_tf}, it might appear that the results at the optimal evolution time are the same for the WCL and SCL case.  This is a result we expect to hold only for the single qubit case.  As we increase the number of qubits $N$, the decoherence time scales with $N$ in such a way that we would not expect AQC to be possible at all in the case of the SCL. To see this, we start from the general SCL master equation~\eqref{eqt:SCL} and assume from now on that the the system operators $\{A_{\alpha} \}$ are single-qubit operators, i.e., that the index $\alpha\in\{1,\dots,N\}$ enumerates the qubits. Our pessimistic conclusions about the SCL will not be improved by considering the case of general decoherence (e.g., with different Pauli operators acting on a given qubit), so this assumption will serve to illustrate the limitations of AQC in the SCL. Moreover, we may assume without loss of generality that the $\{A_{\alpha} \}$ are Hermitian. In this case the $\{A_{\alpha} \}$ operators commute and there exists a mutual diagonalizing basis $\{\ket{a}\}$. In this basis we can write $A_{\alpha}\ket{a} = A_{\alpha a}\ket{a}$, where $A_{\alpha a} \equiv \bra{a} A_{\alpha} \ket{a}$ (the $a$th eigenvalue of $A_\alpha$), and $\rho_{ab} = \bra{a} \rho(t) \ket{b}$. We consider the contribution of the dissipative part and obtain
\bes
\begin{align}
& \bra{a} \mathcal{L}_{\textrm{SCL}}[\rho(t)] \ket{b}  \\
& \quad = \sum_{\alpha, \beta} \gamma_{\alpha \beta}(0) \left(A_{\beta a} A_{\alpha b} - \frac{1}{2} A_{\alpha a} A_{\beta a} - \frac{1}{2} A_{\alpha b} A_{\beta b} \right) \rho_{ab} \notag \\
& \quad= -\frac{1}{2} \sum_{\alpha, \beta}  \gamma_{\alpha \beta}(0) \left( A_{\alpha a} - A_{\alpha b} \right) \left( A_{\beta a} - A_{\beta b} \right) \rho_{ab}
\end{align}
\ees
where we have used $\gamma_{\alpha \beta}(0) = \gamma_{\beta \alpha}(0)$ [which follows from the KMS condition \eqref{eq:KMS}].  Let us now consider two opposite extremes of decoherence.

\subsubsection{Independent Decoherence}

In the case of identical, independent baths, we have $\gamma_{\alpha \beta}(0) = \delta_{\alpha \beta} \gamma(0)$. Thus
\beq
 \bra{a} \mathcal{L}_{\textrm{SCL}}[\rho(t)] \ket{b} =  -r_{ab} \rho_{ab}\ ,
\eeq
where
\beq
\label{eqt:SCLT2}
r_{ab} = \frac{1}{T_2^{(c)}} \sum_{\alpha=1}^N\left[ (A_{\alpha a} - A_{\alpha b})/2 \right]^2
\eeq
is the decay rate of $\rho_{ab}$ in the diagonalizing basis of $A_{\alpha}$, and where $T_2^{(c)} = 1/[2\gamma(0)]$ is the single-qubit dephasing time [Eq.~\eqref{eqt:T2Z}]. 

When the $\{A_{\alpha}\}$ are Pauli operators the eigenvalues are $\pm 1$, so that $A_{\alpha a} = (-1)^{a_\alpha}$, where $a_\alpha \in \{0,1\}$. Thus $|A_{\alpha a} - A_{\alpha b}| = 2d_{\alpha,ab}$ where $d_{\alpha,ab}\in\{0,1\}$ is the Hamming distance between $\ket{a}$ and $\ket{b}$ on the $\alpha$th qubit, and hence 
\beq
0  \leq  r_{ab} = \frac{1}{T_2^{(c)}}  \sum_{\alpha} d_{\alpha,ab}^2 \leq \frac{N}{T_2^{(c)}} \ .
\label{eq:rab}
\eeq
Combining Eqs.~\eqref{eqt:SCLT2} and \eqref{eq:rab} we see that $r_{aa}=0$, i.e., the dissipative part does not affect the populations directly, and all off-diagonal elements decay exponentially with a rate $r_{ab} >0$, which can be as high as $N$ times the single-qubit dephasing rate. 

While we have not explicitly demonstrated that the populations equalize in this setting (as we saw in the single-qubit case in Sec.~\ref{sec:SCL2}), it is still clear that no useful AQC can take place: the ground state of $H_S(t)$ will, in general, be a coherent superposition of the complete set of eigenstates of the $\{ A_\alpha \}$ operators, and we have demonstrated that this superposition decays on multiple timescales, varying from the single-qubit dephasing time $T_2^{(c)}$ to $N$ times this timescale. Thus, to be able to perform useful AQC in the SCL under the independent decoherence model, one must invoke some form of error correction, suppression, or avoidance \cite{Lidar-Brun:book,jordan2006error,PhysRevLett.100.160506,PhysRevLett.108.080501,PAL:13,Young:13,Young:2013fk,Sarovar:2013kx,Ganti:13,Bookatz:2014uq}.

\subsubsection{Collective Decoherence}
In the case of collective decoherence there is only one system operator: $A = \sum_{\alpha=1}^N A_\alpha$, and a single rate $\gamma_{\alpha\beta}(0) \equiv \gamma(0)$. In the diagonalizing basis we have
\begin{align}
 \bra{a} \mathcal{L}_{\textrm{SCL}}[\rho(t)] \ket{b}
& = -\frac{1}{T_2^{(c)}} \left[ (A_{a} - A_{b})/2 \right]^2  \rho_{ab} \ .
\end{align}
%
The eigenvalues are $A_a = N-2h_a$ where $h_a \in \{0,1,\dots N\}$ is the Hamming weight of $\ket{a}$, with multiplicity $\lambda_a = \binom{N}{h_a}$. The ``singlet" case $A_a=0$ (which arises only when $N$ is even) is the well-known decoherence-free subspace (DFS) for collective dephasing, for which $\bra{a} \mathcal{L}_{\textrm{SCL}}[\rho(t)] \ket{b}=0$, and we see that the DFS is spanned by the $\binom{N}{N/2}$ computational basis states having an equal number of $0$'s and $1$'s \cite{Lidar-Brun:book}. Other subspaces, defined by the degeneracy condition $A_a=A_b$, are also decoherence-free, but the latter is the largest such subspace. States belonging to different such subspaces (i.e., for which $h_{a} \neq  h_{b}$) have a positive dephasing rate $\left( h_{a} - h_{b} \right)^2/{T_2^{(c)}}$. However, provided the adiabatic quantum computation is initialized inside a given DFS, it will proceed in a completely unitary manner, and its success probability will be determined purely by the adiabatic condition for closed systems. Thus, under collective dephasing conditions it is possible to support adiabatic quantum computation even subject to the SCL. 

It is not difficult to extend this analysis to the non-commutative case of collective decoherence with different Pauli operators \cite{Zanardi:97c,Lidar:1998fk}, but collectiveness is, of course, a very strong condition (though it can be achieved using dynamical decoupling \cite{Wu:2002zr,Byrd:2002:047901}), so we shall not pursue this further here. 
As we shall see next, in the WCL case the prospects for AQC are significantly more favorable than in the SCL, precisely in the opposite limit of an absence of any degeneracy-inducing symmetries.

\subsection{WCL}
%
\subsubsection{Coherence}
We showed that in the single qubit case, the WCL implies dephasing in the instantaneous energy eigenbasis.  We now wish to check whether this remains true in the multi-qubit case, and whether AQC remains viable in the WCL.
We focus on the dissipative part of the WCL master equation~\eqref{eqt:ME2} and suppress the explicit time-dependence for notational simplicity:
\begin{align} 
\label{eqt:rhoAB}
\mathcal{L}_{\textrm{WCL}}[\rho] &= \sum_{\omega} \sum_{\alpha,\beta} \gamma_{\alpha \beta}(\omega) \times \notag \\
& \qquad \left( L_{\beta,\omega} \rho L_{\alpha,\omega}^{\dagger} - \frac{1}{2} \left\{ L_{\alpha,\omega}^{\dagger} L_{\beta,\omega}, \rho \right\} \right) \ .
\end{align}
We consider the off-diagonal elements of this operator:
\begin{eqnarray} 
\label{eqt:offdiagonal}
\bra{a} \mathcal{L}_{\textrm{WCL}}[\rho] \ket{b} &=& \sum_{\omega} \sum_{\alpha, \beta} \sum_{c,d} \gamma_{\alpha\beta}(\omega) \left[ L_{\beta,\omega, ac} \rho_{cd} L_{\alpha,\omega, db}^{\dagger} \right. \nonumber \\
&& \hspace{-1.75cm} \left.- \frac{1}{2} \left(  \rho_{ac} L_{\alpha,\omega, cd}^{\dagger} L_{\beta,\omega, db} + L_{\alpha,\omega,ac}^{\dagger} L_{\beta,\omega,cd} \rho_{db} \right) \right]
\end{eqnarray}
where now the complete set $\{\ket{a}\}$ is the energy eigenbasis, i.e., the instantaneous  eigenstates of $H_S(t)$, and where $L_{\alpha,\omega,ac} \equiv \bra{a} L_{\alpha,\omega} \ket{c}$. 
We would like to extract the $T_2^{(e)}$ time from this expression. This is not possible in general, but as we show in detail in Appendix \ref{app:CalculateT2}, it is possible under the assumption that there are no accidental symmetries, i.e., that the spectrum is non-degenerate, and that moreover the energy gaps are also non-degenerate, i.e., $\delta_{\eps_{a'} - \eps_a, \eps_{b'} - \eps_b}  = \delta_{a,a'} \delta_{b,b'}$. This allows us to explicitly write the $T_2$ time associated with any pair of energy eigenstates $\ket{a}$ and $\ket{b}$ (with $a\neq b$):
\begin{eqnarray} 
\label{eqt:T2-multiqubit}
\frac{1}{T_2^{(e)}(a,b)}  &=& \nonumber \\ && \hspace{-0.5cm} \frac{1}{2} \sum_{\alpha,\beta} \gamma_{\alpha \beta}(0) \left( A_{\alpha, aa} - A_{\alpha, bb} \right) \left( A_{\beta, aa} - A_{\beta, bb} \right)  \nonumber \\
&&  \hspace{-0.5cm}+ \frac{1}{2} \sum_{\alpha,\beta} \left( \sum_{b'\neq a}  \gamma_{\alpha \beta}(\eps_a - \eps_{b'}) A_{\alpha, a b'} A_{\beta b' a} \right. \nonumber \\
&&  \hspace{-0.5cm} \left. +  \sum_{a'\neq b} \gamma_{\alpha \beta} (\eps_b - \eps_{a'}) A_{\alpha, b a'} A_{\beta, a' b} \right)  \ .
\end{eqnarray}
The single qubit result in Eq.~\eqref{eqt:Sigma} is a special case of Eq.~\eqref{eqt:T2-multiqubit}, as can be seen by taking $a = \eps_-, b = \eps_+$ [from Eq.~\eqref{eqt:1qubiteigenstates}] and noting that $A_{++} = - A_{--} = \frac{s \Gamma}{\lambda(s)}$ and $A_{-+} = A_{+-} = \frac{1-s}{\lambda}$.

In the case of identical, independent baths, we can use $\gamma_{\alpha\beta}(\omega) = \gamma(\omega) \delta_{\alpha\beta}$ to further simplify this expression to:
\begin{eqnarray} 
\label{eqt:T2-multiqubit-simple}
\frac{1}{T_2^{(e)}(a,b)} &=& \frac{1}{T_2^{(c)}}  \sum_{\alpha} \left[( A_{\alpha, aa} - A_{\alpha, bb})/2 \right]^2 \nonumber \\
&& + \frac{1}{2} \sum_{\alpha} \left( \sum_{b'\neq a}  \gamma(\eps_a - \eps_{b'}) |A_{\alpha, a b'}|^2 \right. \nonumber \\
&& \left. +   \sum_{a'\neq b} \gamma(\eps_b - \eps_{a'}) |A_{\alpha, b a'}|^2 \right)  \ ,
\end{eqnarray}
where we have introduced the single-qubit $T_2^{(c)}$ time via $\gamma(0) = 1/[2T_2^{(c)}]$. Since each term in the sums is now manifestly positive, this shows explicitly how a small single qubit $T_2^{(c)}$ time enforces a small $T_2^{(e)}$ dephasing time, but in the instantaneous energy eigenbasis.  

\subsubsection{Ground State Population}

While a small $T_2^{(c)}$ time causes rapid decoherence between energy eigenstates, it does not necessarily translate to a small thermalization time. To see this, starting again from Eq.~\eqref{eqt:rhoAB}, we can determine the rate equations for the populations in the energy eigenbasis.  In the absence of degeneracies (see Appendix \ref{app:CalculateT1} for details), we find that $\dot{\rho}_{00} = \bra{\eps_0} \mathcal{L}_{\textrm{WCL}}[\rho] \ket{\eps_0} = -r_0 \rho_{00} + \sum_{b>0}$, where the sum over $b$ is a relaxation term that repopulates the ground state, and the rate $r_0$ of depopulation of the ground state due to the dissipative dynamics (again for identical independent baths in the absence of degeneracies) is given by:
\beq
r_0 = \sum_{a>0} \gamma(\eps_a -\eps_0) e^{- \beta (\eps_a -\eps_0)} \sum_{\alpha}  |A_{\alpha,0a}|^2\ .
\label{eq:r0}
\eeq
As is evident, the $\gamma(0)$ term is absent in this sum, indicating that the single qubit $T_2^{(c)}$ does not play a detrimental role in depopulating the ground state. This conclusion is robust even in the presence of degeneracies (which we have ignored), since the corresponding $\gamma(0)$ terms would arise due to population transfer between degenerate ground states. A problem would arise in that case only if a degenerate ground state became an excited state later in the evolution.  

We may thus conclude that AQC in the WCL is largely unaffected by a small single-qubit $T_2^{(c)}$: decoherence is between energy eigenstates, which is harmless, and ground state depopulation does not depend on $T_2^{(c)}$, as long the energy gap does not close (a scenario that is detrimental to AQC even in the closed system case). Ground state depopulation is protected by the gap via the Boltzmann factors $e^{- \beta (\eps_a -\eps_0)}$.

\section{SQA-EB: Simulated quantum annealing with an explicit bath}
\label{sec:VIII}
%

Our master equation analysis has allowed us to study two limits: the WCL, where the system-bath interaction is weak relative to the system Hamiltonian, and the SCL, where the system-bath interaction is strong relative to the system Hamiltonian.  In either case, the master equation approach becomes computationally prohibitive when the system size becomes large, since in principle it scales with the square of the dimension of the system Hilbert space.  In practice we are restricted to simulating up to about $15$ qubits in this manner. Furthermore, in order to interpolate between the two limits, an explicit treatment of the bath degrees of freedom is necessary.  In the case of bosonic baths with dephasing interactions, this can be achieved by integrating out the bath degrees of freedom.  For the case of a single qubit, analytic expression for the dynamics of the two-level system can then be found using the ``non-interacting blip approximation'' \cite{RevModPhys.59.1,PhysRevA.35.1436}, and this method can be successful in capturing the dynamics of multi-qubit open system adiabatic quantum computing if the dynamics is effectively restricted to only two levels \cite{Boixo:2014yu}. 

In this section we present a different approach, that is numerically efficient in the sense that its computational cost scales in the same manner as classical Monte Carlo methods. Moreover, this will allow us to probe the intermediate regime, between the WCL and the SCL. The price to be paid is that instead of simulating the dynamics we will be sampling from the instantaneous Gibbs distribution of the system.

To be explicit we now restrict our attention to time-dependent system Hamiltonians of the form of the transverse Ising model,
\beq \label{eq:TransverseIsing}
H_S(t) = -A(t) \sum_i \sigma^x_i + B(t) \left[ \sum_{i} h_i \sigma^z_i + \sum_{i<j} J_{ij} \sigma_i^z \sigma_j^z \right] \ .
\eeq
In this setting adiabatic quantum computing is also known as quantum annealing \cite{kadowaki_quantum_1998,morita:125210}.  In simulated quantum annealing \cite{sqa1,Santoro,q108} (SQA), Monte Carlo dynamics are used to sample from the instantaneous Gibbs state along the annealing evolution.  For example, in the case of discrete-time Monte Carlo, for each fixed time $t$ in Eq.~\eqref{eq:TransverseIsing}, Monte Carlo sampling of the thermal state associated with the transverse Ising Hamiltonian is done by sampling the dual classical spin system with:
\begin{eqnarray} \label{eq:action}
\beta \mathcal{H}_S(t) &=& \frac{\beta}{N_{\tau}} B(t) \sum_\tau \left[ \sum_{i} h_i \mu_{i,\tau} + \sum_{i<j} J_{ij} \mu_{i,\tau} \mu_{j,\tau} \right] \nonumber \\
&&-J_{\perp}(t)  \sum_{i,\tau} \mu_{i,\tau} \mu_{i,\tau+1} \ ,
\end{eqnarray}
where $\beta$ is the inverse temperature of the Monte Carlo simulation, $N_\tau$ is the number of Trotter slices used along the Trotter direction (also referred to as the imaginary-time or time-like direction), $\mu_{i,\tau}$ denotes the $i$th classical spin on the $\tau$th Trotter slice, and 
\beq
J_{\perp}(t) \equiv -  \frac{1}{2}  \ln(\tanh( \beta A(t) /N_{\tau})) > 0
\eeq 
is the nearest-neighbor coupling strength along the Trotter direction.  Although this does not capture the unitary dynamics of the quantum system, the sampling of the instantaneous Gibbs state mimics the thermalization process towards the Gibbs state in the WCL master equation, with the advantage that the simulation remains efficient, so that the system size can be made large.  However, even as the temperature is increased, neither the WCL master equation nor the standard SQA methods capture the decoherence of the qubits into classical bits, unlike the SCL master equation.  

\begin{figure}[t] 
   \centering
   \includegraphics[width=0.5\columnwidth]{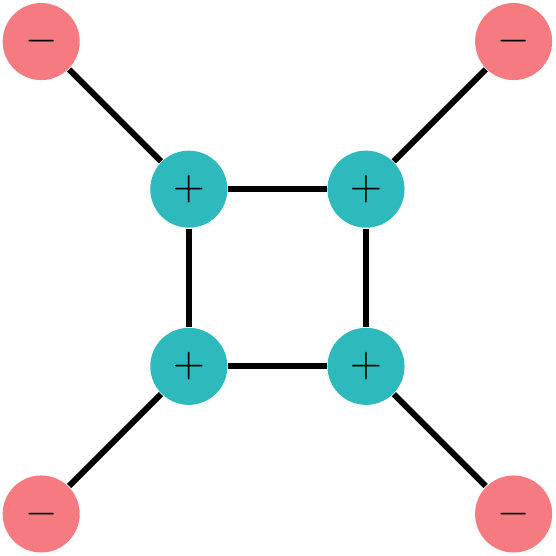} 
   \caption{The 8 qubit ``quantum signature'' Hamiltonian $H_{\mathrm{Ising}}$ studied in Ref.~\cite{q-sig}.  The spins are depicted by colored disks.  All spin-spin couplings (black lines connecting spins) are ferromagnetic with magnitude $1$ and the signs of the local fields (of magnitude $1$) are indicated within the disks.  We use the sign convention $H_{\mathrm{Ising}} = - \sum_i h_i \sigma_i^z + \sum_{i<j} J_{ij} \sigma_i^z \sigma_j^z$.}
   \label{fig:QSig}
\end{figure}
\begin{figure*}[t] 
	\subfigure[\ $100$ sweeps ]{ \includegraphics[width=0.3\textwidth]{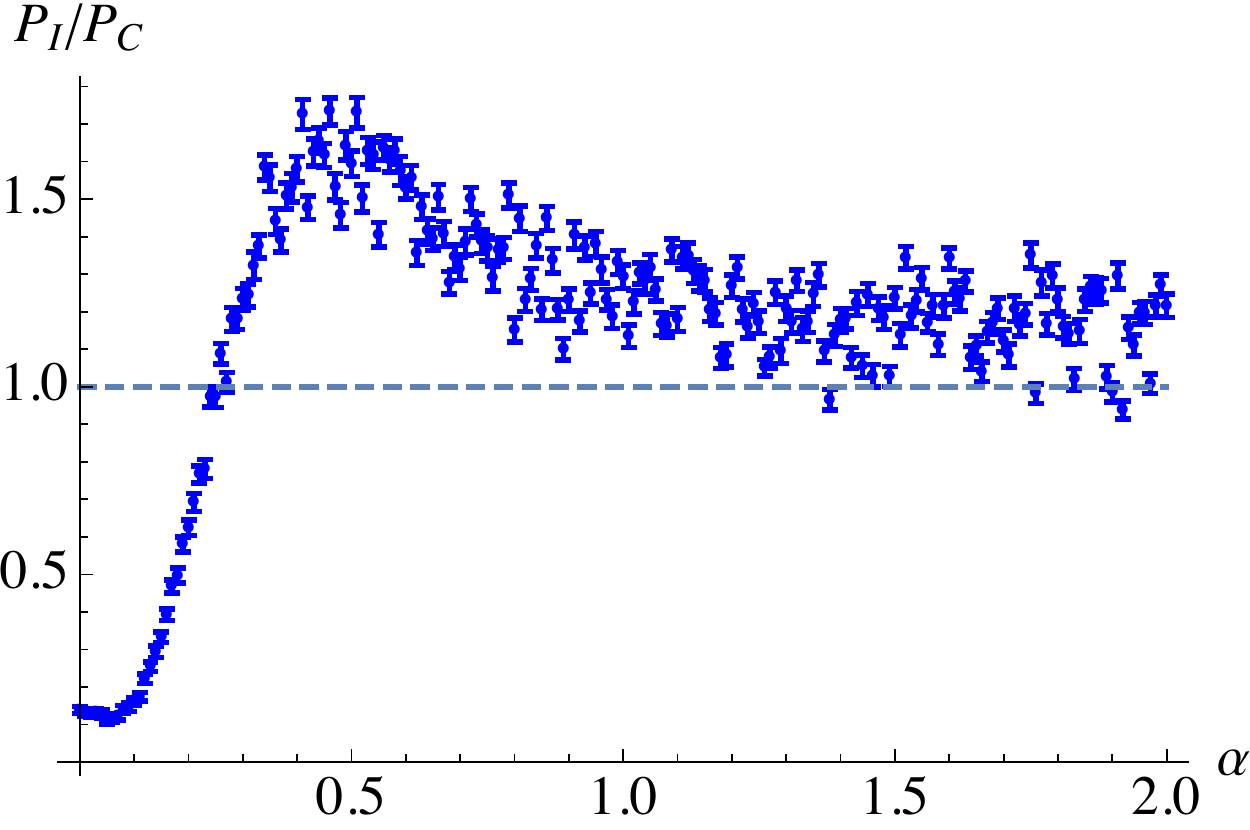} \label{fig:SQA_100}}
	\subfigure[\ $200$ sweeps ]{ \includegraphics[width=0.3\textwidth]{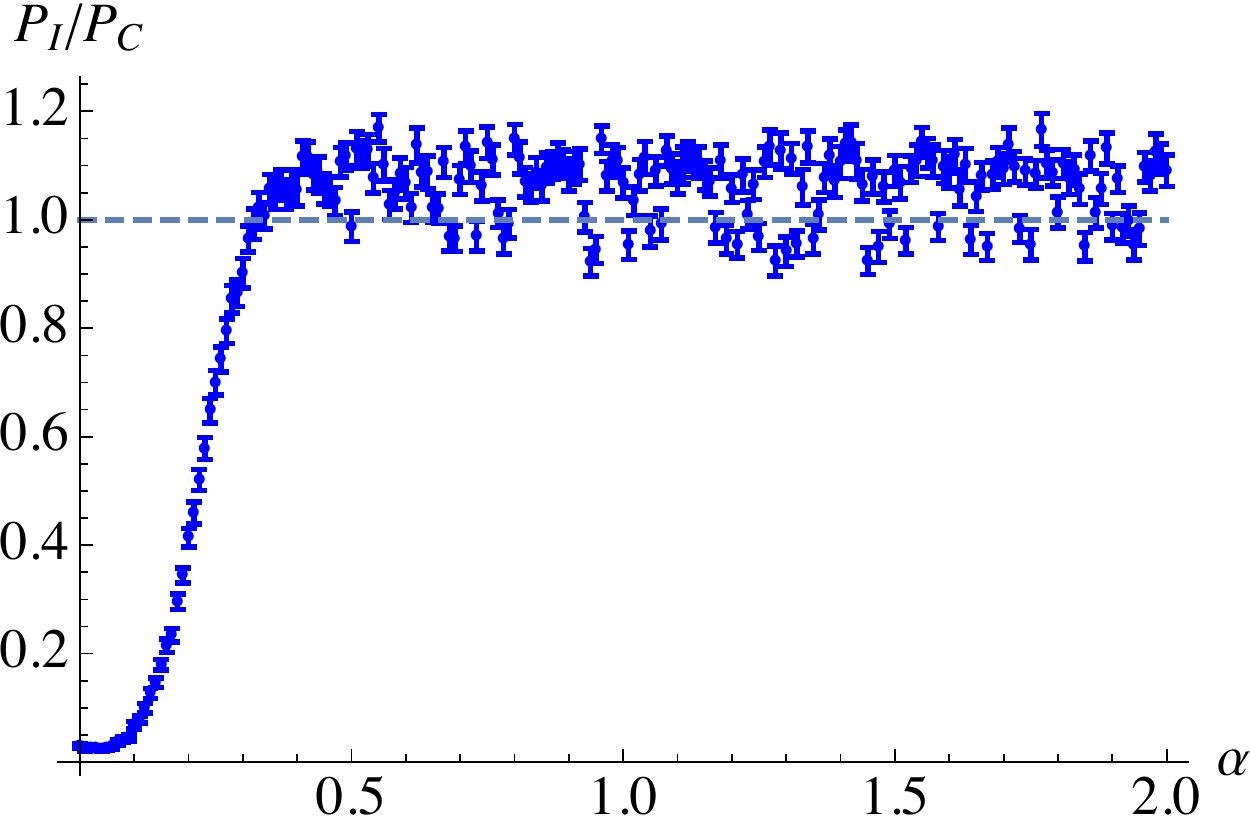} \label{fig:SQA_200}}
	\subfigure[\ $500$ sweeps]{ \includegraphics[width=0.3\textwidth]{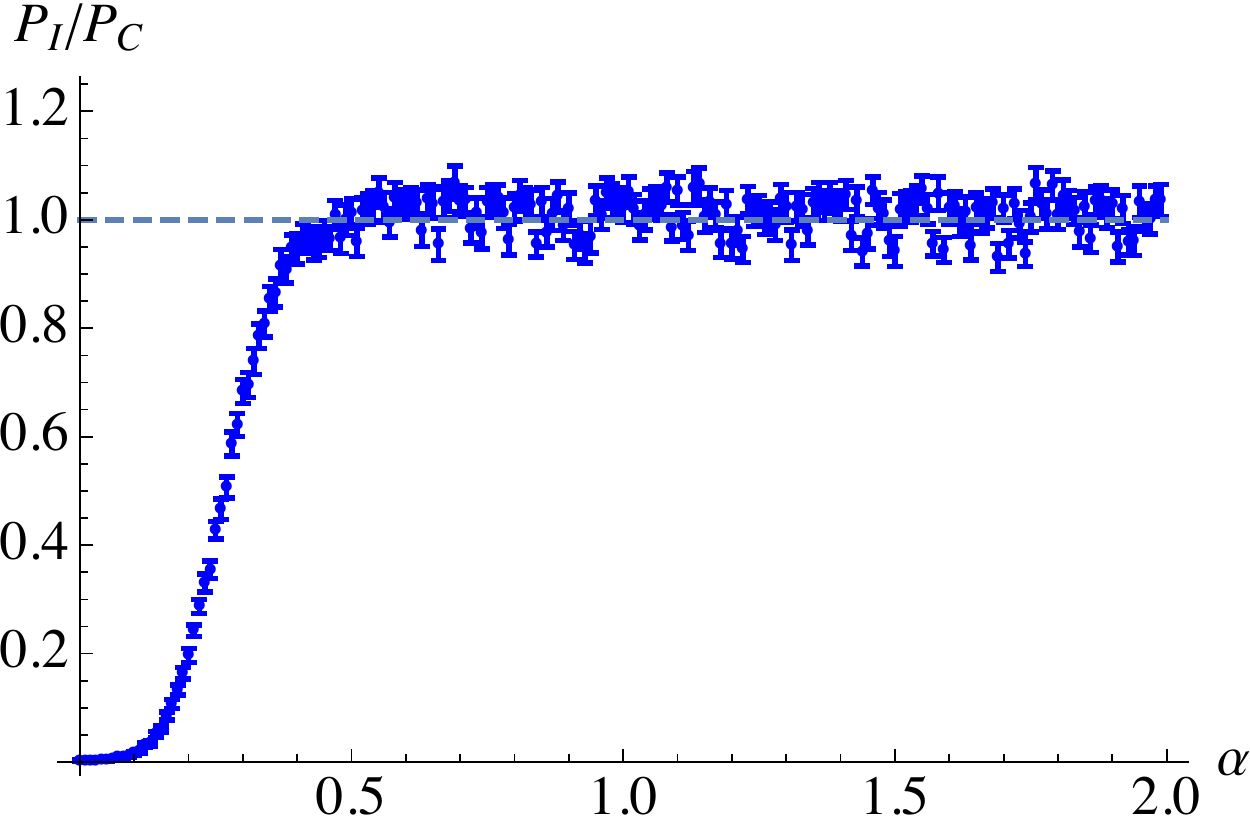} \label{fig:SQA_500}}
   	\caption{Simulation results for the final-time ($t=t_f$) ratio $P_{\mathrm{I}}/P_{\mathrm{C}}$ using SQA with explicit baths of varying system-bath interaction strengths $\alpha$, as defined in Eq.~\eqref{eq:SQA_bath}.  For all sweep values, the ratio grows as $\alpha$ is increased, eventually crossing or saturating at $P_{\mathrm{I}}/P_{\mathrm{C}} = 1$, denoted by the dashed line.  SQA simulation parameters: $\beta = 10$, $N_{\tau} = 64$, a linear annealing schedule ($A(s) = 1-s$, $B(s) = s$), and a total of (a) $100$, (b) $200$, and (c) $500$ sweeps.  $10^5$ SQA runs were performed for each $\alpha$ value. Error bars were generated by performing $100$ bootstraps on the $10^5$ runs and taking twice the standard deviation of the $P_{\mathrm{I}}/P_{\mathrm{C}}$ values.}
   \label{fig:SQA}
\end{figure*}

In this section we show that by incorporating an explicit bath in SQA, we can interpolate between SQA and classical simulated annealing (SA), where the qubits have fully decohered into classical bits that undergo bit flip updates. We call this method SQA-EB (EB for ``explicit bath"). In addition to Eq.~\eqref{eq:TransverseIsing}, we include independent yet identical bosonic baths for each qubit with a dephasing system-bath interaction:
\beq
 H_{\mathrm{SB}} = \sum_{i=1}^{N} \sigma^z_i \otimes \sum_k g_k \left(b_{i,k} + b_{i,k}^{\dagger} \right) \ .
 \eeq
Assuming the baths to have an Ohmic spectral function, we can analytically integrate out the bosonic degrees of freedom \cite{PhysRevB.9.215,RevModPhys.59.1}, and the standard discrete-time quantum Monte Carlo action of Eq.~\eqref{eq:action} is supplemented with the additional term \cite{PhysRevLett.94.047201,Werner:2005fb}:
\beq \label{eq:SQA_bath}
\beta \mathcal{H}_{\mathrm{SB}} = - \alpha \sum_{i=1}^{N} \sum_{\tau =1}^{N_{\tau}} \sum_{\tau' =\tau+1}^{N_{\tau}} \frac{\mu_{i,\tau} \mu_{i, \tau'}}{\sin^2 \left( \frac{\pi}{N_{\tau}}|\tau - \tau'| \right)} \ ,
\eeq
where $\alpha>0$ is the system-bath strength.  This term introduces (non-local) ferromagnetic couplings between all spins along the Trotter direction of the dual classical spin system. If this coupling overwhelms the transverse field term in the action, which is the only other source of couplings in the Trotter direction for the transverse Ising Hamiltonian, then it is conceivable that all the spins in the imaginary-time direction behave as one large spin flipping together.  In this sense, updates in the dual classical spin system should behave as SA updates on the Ising part of the Hamiltonian.

In order to test this intuition, we use the $8$-qubit ``quantum signature'' Hamiltonian proposed in ref.~\cite{q-sig}, depicted in Fig.~\ref{fig:QSig}.  This Ising Hamiltonian has the feature of having a $17$-fold degenerate ground state, $16$ of which are connected via single spin flips, so they are called the ``cluster'' ground state.  The remaining ground state is at least $4$ spin flips away, so is referred to as the ``isolated'' ground state.  We  denote the average population in the $16$ cluster ground states by $P_{\mathrm{C}}$ and the population in the isolated ground state by $P_{I}$, both at the final time $t=t_f$. In the thermal equilibrium state all ground states are equally probable, i.e., $P_\mathrm{I}/P_{\mathrm{C}} = 1$. In Ref.~\cite{q-sig} it was shown that SA and quantum annealing can be differentiated by the value of the ratio $P_\mathrm{I}/P_\mathrm{C}$: SA will always have $P_\mathrm{I}/P_\mathrm{C} \geq 1$, i.e., SA preferentially populates the isolated ground state relative to any given cluster ground state, while quantum annealing will typically have $P_\mathrm{I}/P_\mathrm{C} < 1$, i.e., quantum annealing preferentially populates the cluster ground states. These differences can be understood by studying the corresponding classical and quantum spectra.  The classical spectrum is such that any random state can reach the ground states without encountering any local minima, and there are more paths to reach the isolated ground state than any given cluster ground state.  This explains why SA favors the isolated ground state. Using first order perturbation theory, the degeneracy of the ground states is broken by the introduction of the transverse field, and the isolated state has no overlap with the perturbed ground state.  Hence, quantum annealing will generically not populate the isolated ground state.  More details can be found in Refs.~\cite{q-sig,q-sig2}.

If SQA with an explicit bath can interpolate between QA and SA by tuning the system-bath strength, it should be able to interpolate between the two regimes of $P_\mathrm{I}/P_\mathrm{C} < 1$ and $P_\mathrm{I}/P_\mathrm{C} > 1$ by increasing the system-bath coupling strength.  Our simulations validate this picture, as shown in Fig.~\ref{fig:SQA}.  Increasing the system-bath strength raises the $P_\mathrm{I}/P_{\mathrm{C}}$ ratio from close to $0$ to above $1$, thus causing the final distribution of ground states to increasingly resemble the classical limit. The intermediate coupling strength regime is the one that cannot be captured by the WCL. 

We further test the dependence on the total evolution time, measured in terms of the the number of sweeps, where a sweep is one complete Monte Carlo update of all spins.
In the case of a small number of sweeps [see Fig.~\ref{fig:SQA_100}], we observe the ratio first peaking and then decreasing towards $1$. This shows, as expected, that for short evolution times the system requires a strong system-bath coupling in order to reach its equilibrium state (with $P_\mathrm{I}/P_{\mathrm{C}} = 1$), i.e., that the system equilibrates faster as the system-bath strength is increased. When the number of sweeps is high [see Figs.~\ref{fig:SQA_200} and \ref{fig:SQA_500}], the non-monotonic behavior is replaced with a monotonic approach towards the thermal equilibrium value of $P_\mathrm{I}/P_{\mathrm{C}} =1$.  In all cases shown here, the ground state probability is close to one.

We can understand the WCL as being able to capture the $P_\mathrm{I}/P_{\mathrm{C}} \leq 1$ regime. Values of $P_\mathrm{I}/P_{\mathrm{C}}$ away from this, as seen in Fig.~\ref{fig:SQA}, represent regimes that cannot be captured by the WCL, and are thus of particular interest for SQA methods that include an explicit bath dependence, as done here.

%
\section{Conclusions}
\label{sec:IX}
%
In this work we revisited the problem of decoherence in adiabatic quantum computation (AQC), using a master equation approach and a quantum Monte Carlo approach. We argued that the common perception that decoherence is always detrimental for quantum computation should be qualified in the adiabatic case, since the extent of decoherence-induced damage depends on whether the adiabatic quantum computer operates in the weak or singular coupling limit (WCL or SCL), or perhaps in an intermediate regime. A well-engineered device should operate in the WCL, whence 
dephasing occurs in the system's instantaneous energy eigenbasis. This is a form of decoherence that preserves the coherence of the instantaneous quantum ground state (and all other instantaneous energy eigenstates). Therefore this form of decoherence, even if extremely rapid, does not necessarily negatively impact AQC. This is in stark contrast to dephasing in the circuit-paradigm of quantum computing, where dephasing in the computational basis spoils the efficiency of the computation and mandates error correction.  Of course, there is no free lunch in the AQC either, as the WCL does allow decoherence in the form of thermal excitations, which can depopulate the ground state and thus negatively impact AQC. However, the final state reached in the long-evolution time limit is the thermal Gibbs state, and as long as the ground state population does not decrease with increasing system size, AQC can succeed. Whether a quantum speedup is possible is a separate question that depends on how rapidly the Gibbs state is reached, and which we have not addressed in this work. 

A particularly noteworthy result we have demonstrated here (see Fig.~\ref{fig:1Qubit_tf}) is that there is an optimal (problem-dependent) evolution time $t_f$ that is much shorter than the adiabatic time-scale, where the ground state population can be significantly higher than that in the thermal Gibbs state. We have also presented evidence that the optimal $t_f$ decreases with the strength of the system-bath coupling (see Fig.~\ref{1Qubit_optimal}). This shows that it can be advantageous to run AQC with a much shorter duration total evolution than that suggested by the standard (heuristic) inverse gap criterion. Finding this optimal evolution time might be done, e.g., by picking some initial $t_f$, finding the ground state probability, then repeating the experiment using both $t_f/2$ and $2t_f$, etc., thus performing a grid search that will converge rapidly on the optimal $t_f$. Another interesting possibility is to apply the theory of optimal stopping times for continuous processes \cite{book:optimal-stopping}.

It is also interesting to note that the standard strategy of suppressing detrimental thermal excitations by increasing the system's energy gap is only guaranteed to work in the very large gap limit, since as we have demonstrated using an Ohmic bath model, the excitation rate first rises and only then decreases as a function of the energy gap. This is true also in the multi-qubit setting [Eq.~\eqref{eq:r0}], since the suppression of thermal excitations via the gap-dependent Boltzmann factor is counteracted by the gap-dependent excitation rate $\gamma$. These conclusions apply under the assumption that the Markovian adiabatic master equation we have used here holds; the situation in the non-Markovian setting can be different, and even an infinite energy gap may not suffice \cite{Marvian:2014nr}.

In stark contrast to the WCL results, in the case of the SCL decoherence occurs in the computational basis, resulting in  the loss of instantaneous quantum ground state coherence. AQC then becomes impossible since the resulting final state is essentially fully mixed, i.e., the ground state population drops exponentially with system size.

The WCL and the SCL describe two dynamical limits of the system-bath coupling, and we have described an approach (SQA-EB) that applies in the intermediate coupling regime. This is accomplished by integrating out the bath degrees of freedom in the path integral formalism, resulting in an effective action for the spin system, that can be used with quantum Monte Carlo methods to sample from the instantaneous Gibbs distribution.  A method such as SQA-EB for simulating open system quantum annealing beyond the WCL and the SCL is particularly useful in studying the effect of decoherence, as it provides us with a mechanism to controllably decohere qubits into classical bits.  Although SQA-EB does not capture unitary dynamics during the quantum evolution, it has the advantage of enabling the study of large system sizes.  This provides us with a valuable tool for modeling large open quantum annealing systems. It is particularly important as a means to go beyond the WCL, whose validity depends on the system's energy gap remaining large compared to the system-bath coupling strength [Eq.~\eqref{eq:8a}], an assumption that will be violated for sufficiently large systems encoding computationally interesting ground states, whose gaps are known to shrink rapidly as a function of system size. Methods such as SQA-EB will play an important role in deciding the relative advantage of quantum annealing over classical annealing \cite{Nishimori:2015dp}. An outstanding open problem is to develop a method that generalizes the WCL and SCL adiabatic master equations and can capture adiabatic quantum dynamics in the full intermediate coupling regime.

\acknowledgements
The authors thank Matthias Troyer for useful discussions and references.  The computing resources were provided by the USC Center for High Performance Computing and Communications.  The authors acknowledge support under ARO grant number W911NF-12-1-0523, ARO MURI Grant No. W911NF-11-1-0268.

%

\appendix
\section{Off-diagonal components and decoherence in the WCL} 
\label{app:CalculateT2}

We would like to isolate the contribution of the diagonal elements of $\rho$ to the off-diagonal term $\bra{a} \mathcal{L}_{\textrm{WCL}}[\rho] \ket{b} $ given in Eq.~\eqref{eqt:offdiagonal}. The reason is that we would like to demonstrate that under the right assumptions (non-degeneracy, as explained below), the diagonal elements of $\rho$ do not appear in this off-diagonal term, which means that the off-diagonal elements of $\rho$ evolve independently from its diagonal elements. This will allow us to extract the $T_2$ time. 

In addition to the KMS condition $\gamma_{\alpha \beta} (-\omega) = e^{-\upbeta \omega} \gamma_{\beta \alpha}(\omega)$ we repeatedly use the following easily verified identities:
\bes
\begin{align}
\label{eq:LL1}
& L_{\alpha, -\omega} = L_{\alpha,\omega}^{\dagger} \\
\label{eq:LL2}
& L_{\alpha, \omega\neq 0, aa} = 0 \\
\label{eq:LL3}
& L_{\alpha,\omega,ab} = L^\dagger_{\alpha,\omega,ba} = \delta_{\omega,\eps_b-\eps_a} A_{\alpha,ab}\\
\label{eq:LL4}
& L_{\beta,\omega, ac} L_{\alpha,\omega,cb}^\dagger , L_{\beta,\omega, ac}^\dagger L_{\alpha,\omega,cb} \propto \delta_{\eps_a,\eps_b} \ .
\end{align}
\ees
Eq.~\eqref{eq:LL1} follows from Eq.~\eqref{eq:Lindblad2} provided the $A_\alpha$ operators are Hermitian, which can be assumed without loss of generality. Eqs.~\eqref{eq:LL2} and \eqref{eq:LL3} follow directly by taking matrix elements of $L_{\alpha,\omega}$. Eq.~\eqref{eq:LL4} follow since the term $L_{\beta,\omega, ac}$ is non-zero only if the energies associated with the states $\{\ket{\eps_a}, \ket{\eps_c}\}$ satisfy $\omega = \eps_c - \eps_a$; similarly, the term $L^{\dagger}_{\beta,\omega, cb}$ is non-zero only if $\omega = \eps_c - \eps_b$.  Therefore the term $L_{\beta,\omega, ac} L_{\alpha,\omega,cb}^\dagger$ is non-zero only if $\eps_a = \eps_b$, i.e., the states are degenerate.  

For convenience we reproduce Eq.~\eqref{eqt:offdiagonal} here in three parts:
\bes
\label{eq:53abc}
\begin{align} 
\label{eq:53a}
& \bra{a} \mathcal{L}_{\textrm{WCL}}[\rho] \ket{b} = P_1 + P_2 +P_3 \\
& P_1 = \sum_{\omega} \sum_{\alpha, \beta} \sum_{c,d} \gamma_{\alpha\beta}(\omega) L_{\beta,\omega, ac} \rho_{cd} L_{\alpha,\omega, db}^{\dagger} \\
\label{eq:53b}
& P_2 =  - \frac{1}{2} \sum_{\omega} \sum_{\alpha, \beta} \sum_{c,d} \gamma_{\alpha\beta}(\omega) \rho_{ac} L_{\alpha,\omega, cd}^{\dagger} L_{\beta,\omega, db} \\
& P_3 = - \frac{1}{2} \sum_{\omega} \sum_{\alpha, \beta} \sum_{c,d} \gamma_{\alpha\beta}(\omega)  L_{\alpha,\omega,ac}^{\dagger} L_{\beta,\omega,cd} \rho_{db} 
\end{align}
\ees

Considering Eq.~\eqref{eq:53abc}, there are three opportunities for the diagonal elements to appear, in $\rho_{cd}$ with $c=d$ ($P_1$), in $\rho_{ac}$ with $c=a$ ($P_2$), and in $\rho_{db}$ with $d=b$ ($P_3$). Thus we consider these separately. 
Setting $c = d$ and $\omega = 0$ in $P_1$ yields $\sum_{\alpha \beta} \gamma_{\alpha \beta}(0) \sum_{c} L_{\beta,0,ac} L_{\alpha,0,cb}^{\dagger} \rho_{cc}$. We split this into $c \neq a,b$, which gives \eqref{eqt:rhocc}; $c=a$, which gives the first term in Eq.~\eqref{eqt:rhoaa}, and $c=b$, which gives the first term in Eq.~\eqref{eqt:rhobb}. 
Setting $c=a$ and $\omega = 0$ in $P_2$ gives the second term in Eq.~\eqref{eqt:rhoaa}. Setting $d=b$ and $\omega = 0$ in $P_3$ gives the second term in Eq.~\eqref{eqt:rhobb}.
\bes 
\label{eqt:diagonalrho}
\begin{align}
& \sum_{\alpha \beta} \gamma_{\alpha \beta}(0) \sum_{c \neq a,b} L_{\beta,0,ac} L_{\alpha,0,cb}^{\dagger} \rho_{cc}  \label{eqt:rhocc} \\
&\sum_{\alpha \beta} \gamma_{\alpha \beta}(0)\left( L_{\beta,0,aa} L_{\alpha,0,ab}^\dagger - \frac{1}{2} \sum_c   L_{\alpha,0,ac}^\dagger L_{\beta,0,cb} \right) \rho_{aa} \label{eqt:rhoaa} \\
&\sum_{\alpha \beta} \gamma_{\alpha \beta}(0)\left( L_{\beta,0,ab} L_{\alpha,0,bb}^\dagger - \frac{1}{2} \sum_c   L_{\alpha,0,ac}^\dagger L_{\beta,0,cb}\right) \rho_{bb}  \label{eqt:rhobb}
 \end{align}
 \ees
We repeat the same process for $\omega \neq 0$ and use the KMS condition, but now all the diagonal terms (with equal Roman subscripts on the $L$ operators) vanish due to Eq.~\eqref{eq:LL2}. With this in mind Eq.~\eqref{eqt:diagonalrho2} can be read off directly from Eq.~\eqref{eqt:diagonalrho}.
\bes 
\label{eqt:diagonalrho2}
\begin{align}
\label{eq:55a}
&\sum_{\omega>0}  \sum_{\alpha \beta} \gamma_{\alpha \beta}(\omega) \sum_{c \neq a,b} \left( L_{\beta,\omega,ac} L_{\alpha,\omega,cb}^{\dagger} \right.  \\
&\hspace{4cm} \left.+ e^{-\upbeta \omega} L_{\alpha,\omega,ac}^{\dagger}  L_{\beta,\omega,cb} \right) \rho_{cc} \nonumber \\
\label{eq:55b}
&- \frac{1}{2}\sum_{\omega>0}  \sum_{\alpha \beta} \gamma_{\alpha \beta}(\omega)  \sum_{c \neq a,b}   \left( L_{\alpha,\omega,ac}^\dagger L_{\beta,\omega,cb}  \right.  \\
&\hspace{4cm} \left.+ e^{-\upbeta \omega} L_{\beta,\omega,ac} L_{\alpha,\omega,cb}^\dagger  \right)  \rho_{aa} \nonumber \\
\label{eq:55c}
&- \frac{1}{2} \sum_{\omega>0}  \sum_{\alpha \beta} \gamma_{\alpha \beta}(\omega )  \sum_{c \neq a,b}   \left( L_{\alpha,\omega,ac}^\dagger L_{\beta,\omega,cb}  \right.  \\
&\hspace{4cm} \left.+ e^{-\upbeta \omega} L_{\beta,\omega,ac} L_{\alpha,\omega,cb}^\dagger  \right)  \rho_{bb} \nonumber
\end{align}
\ees
%
We now make the simplifying assumption that \emph{no two states are degenerate}, i.e., $\eps_a \neq \eps_b$ $\forall a\neq b$.  It then follows from Eq.~\eqref{eq:LL4} that all terms in Eqs.~\eqref{eqt:diagonalrho} and \eqref{eqt:diagonalrho2} vanish since they are composed entirely of products of operators of the type appearing in Eq.~\eqref{eq:LL4}.  

Therefore, we have shown that in the absence of degeneracies the equations for the dissipative dynamics of the off-diagonal density matrix elements do not involve diagonal elements of the density matrix.  We thus consider the contribution of the off-diagonal elements next. To this end we set $c\neq d$ in $P_1$, $c\neq a$ in $P_2$, and $d\neq b$ in $P_3$. This yields, respectively:
\bes 
\begin{align}
& P'_1 = \sum_{\omega} \sum_{\alpha, \beta} \sum_{c\neq d} \gamma_{\alpha\beta}(\omega) L_{\beta,\omega, ac} \rho_{cd} L_{\alpha,\omega, db}^{\dagger} \\
\label{eq:53b}
& P'_2 =  - \frac{1}{2} \sum_{\omega} \sum_{\alpha, \beta} \sum_{c\neq a} \sum_d \gamma_{\alpha\beta}(\omega) \rho_{ac} L_{\alpha,\omega, cd}^{\dagger} L_{\beta,\omega, db} \\
& P'_3 = - \frac{1}{2} \sum_{\omega} \sum_{\alpha, \beta} \sum_c \sum_{d\neq b} \gamma_{\alpha\beta}(\omega)  L_{\alpha,\omega,ac}^{\dagger} L_{\beta,\omega,cd} \rho_{db} 
\end{align}
\ees
Considering $P'_1$, using Eq.~\eqref{eq:LL3} we have $L_{\beta,\omega, ac}L_{\alpha,\omega, db}^{\dagger} = \delta_{\omega,\eps_c-\eps_a}\delta_{\omega,\eps_d-\eps_b}A_{\beta,ac} A_{\alpha,db}$. 
Likewise, considering $P'_2$ we have $L_{\alpha,\omega, cd}^{\dagger} L_{\beta,\omega, db} = \delta_{\omega,\eps_c-\eps_d}\delta_{\omega,\eps_b-\eps_d}A_{\alpha,cd}A_{\beta,db} = \delta_{\omega, \eps_b - \eps_d} \delta_{\eps_b,\eps_c}A_{\alpha,cd}A_{\beta,db}$, and considering $P'_3$ we have 
$L_{\alpha,\omega,ac}^{\dagger} L_{\beta,\omega,cd}   = \delta_{\omega,\eps_a-\eps_c}\delta_{\omega,\eps_d-\eps_c}A_{\alpha,ac}A_{\beta,cd}=\delta_{\omega,\eps_a-\eps_c} \delta_{\eps_a,\eps_d}A_{\alpha,ac}A_{\beta,cd}$.  In the absence of degeneracies, we can further simplify by using that $\delta_{\eps_b,\eps_c} = \delta_{b,c}$ and $\delta_{\eps_a, \eps_d} = \delta_{a,d}$.  This then gives simplified expressions for $P_1'$, $P_2'$, $P_3'$:
\bes
\begin{align}
P_1' = & \sum_{\alpha,\beta} \sum_{c\neq d} \gamma_{\alpha \beta} (\eps_d-\eps_b) \delta_{\eps_c-\eps_a,\eps_d-\eps_b} A_{\beta, ac} A_{\alpha,db} \rho_{cd} \label{eqt:A6a} \\
P_2'  = & - \frac{1}{2} \sum_{\alpha, \beta} \sum_d \gamma_{\alpha \beta} (\eps_b- \eps_d) A_{\alpha, bd} A_{\beta,db} \rho_{ab} \\
P_3' = & - \frac{1}{2} \sum_{\alpha, \beta} \sum_c \gamma_{\alpha \beta} (\eps_a - \eps_c) A_{\alpha, a c} A_{\beta, c a} \rho_{ab} 
\end{align}
\ees

We next consider the contribution of the unitary dynamics, $-i \bra{a}\left[H_S(t) + H_{LS}(t), \rho(t) \right]\ket{b}$ [including the Lamb shift, Eq.~\eqref{eq:H_LS}]. By a similar argument to the one above, it is not hard to see that this too only involves off-diagonal density matrix elements. Then, after combining with $\bra{a} \mathcal{L}_{\textrm{WCL}}[\rho] \ket{b} = P'_1 + P'_2 +P'_3$, we finally obtain for the off-diagonal elements:
\begin{eqnarray} 
\label{eqt:rhoAB2}
\bra{\eps_a(t)} \frac{d}{dt} \rho \ket{\eps_b(t)} & = & - i \left( \eps_a - \eps_b \right) \rho_{ab} \nonumber \\
&& \hspace{-3cm} - i \sum_{\alpha, \beta}\left[  \sum_{b'}S_{\alpha \beta} (\eps_a - \eps_{b'})  A_{\alpha, a b'} A_{\beta, b' a} \rho_{a b} \right. \nonumber \\
&& \hspace{-1.8cm}\left. -  \sum_{a'}S_{\alpha \beta} (\eps_b - \eps_{a'}) A_{\alpha, b a'} A_{\beta, a' b} \rho_{a b}\right] \nonumber \\
&& \hspace{-3cm}+ \sum_{\alpha ,\beta} \left[ \sum_{a' \neq b'} \gamma_{\alpha \beta} (\eps_{b'} - \eps_b) \delta_{\eps_{a'} - \eps_a, \eps_{b'} - \eps_b}  A_{\beta, a a'} A_{\alpha, b' b} \rho_{a' b'} \right. \nonumber \\
&& \hspace{-1.8cm} - \frac{1}{2}   \sum_{a'} \gamma_{\alpha \beta} (\eps_b - \eps_{a'}) A_{\alpha, b a'} A_{\beta, a' b}\rho_{a b} \nonumber \\
&& \hspace{-1.8cm} \left.- \frac{1}{2} \sum_{b'} \gamma_{\alpha \beta}(\eps_a - \eps_{b'}) A_{\alpha, a b'} A_{\beta b' a} \rho_{a b} \right] \ .
\end{eqnarray}
The important point is that only off-diagonal elements of the density matrix appear in this expression. By complete positivity of the Lindblad form of the adiabatic master equation \cite{ABLZ:12-SI}, the instantaneous eigenvalues of this system of linear equations must have magnitude less than or equal to zero. It follows that the off-diagonal elements of the density matrix all decay to zero, and hence that coherence between instantaneous energy eigenbasis states decays away.  (The case of zero eigenvalues correspond to stationary states, which include the instantaneous Gibbs state and any decoherence-free subspace. For the case of independent decoherence, where each qubit interacts with a separate bath, the latter is not possible.)

The system equations~\eqref{eqt:rhoAB2} is coupled due to the appearance of $\rho_{a'b'}$ in the fourth line, and this prevents us from easily extracting the $T_2^{(e)}$ time. This difficulty arises from the $P_1'$ term in Eq.~\eqref{eqt:A6a}. We can make analytic progress in finding the $T_2^{(e)}$ time if further assume that not only the energies are non-degenerate, but also the energy gaps. I.e., we assume that $\delta_{\eps_c-\eps_a,\eps_d-\eps_b} \propto \delta_{a,c}\delta_{b,d}$. Under this assumption $P'_1$ in Eq.~\eqref{eqt:A6a} is proportional to $\gamma_{\alpha\beta}(0)$. After extracting the terms proportional to $\gamma_{\alpha\beta}(0)$ from $P'_2$ and $P'_3$ this then reduces the 
dissipative contribution to the equation for the off-diagonal density matrix elements to:
\bes
\begin{align}
& \bra{a} \mathcal{L}_{\textrm{WCL}}[\rho] \ket{b} =  \sum_{\alpha, \beta}\left[ \gamma_{\alpha \beta}(0) \left(A_{\beta,aa} A_{\alpha,bb} - \frac{1}{2} A_{\alpha, aa} A_{\beta,aa}  \right. \right. \nonumber \\
& \left. \left. - \frac{1}{2} A_{\alpha, bb} A_{\beta,bb}  \right) -  \frac{1}{2}   \sum_{a' \neq b} \gamma_{\alpha \beta} (\eps_b - \eps_{a'}) A_{\alpha, b a'} A_{\beta, a' b}\right. \nonumber \\
& \left. - \frac{1}{2} \sum_{b' \neq a}\gamma_{\alpha \beta}(\eps_a - \eps_{b'}) A_{\alpha, a b'} A_{\beta b' a}   \right]  \rho_{a b}
\end{align}
\ees
The $T_2^{(e)}$ time can now be simply read off to give the expression in Eq.~\eqref{eqt:T2-multiqubit}.

\section{Ground state population loss in the WCL} 
\label{app:CalculateT1}
Starting again from Eq.~\eqref{eqt:rhoAB}, we can write:
\begin{align} 
\label{eqt:rate}
& \bra{a} \mathcal{L}_{\textrm{WCL}}[\rho] \ket{a} = \sum_{\omega} \sum_{\alpha,\beta} \gamma_{\alpha \beta}
(\omega) \times \notag \\
&\quad \sum_{c,d} \left( L_{\beta,\omega,ac} \rho_{cd} L_{\alpha,\omega,da}^{\dagger}
- \frac{1}{2} L_{\alpha,\omega,ac}^{\dagger} L_{\beta,\omega,cd} \rho_{da}  \right.  \nonumber \\ 
& \left. \qquad\qquad
- \frac{1}{2} \rho_{ac} L_{\alpha,\omega,cd}^{\dagger} L_{\beta,\omega,da} \right) \ .
\end{align}
Following a procedure similar to that used in Appendix~\ref{app:CalculateT2}, we note this expression can be significantly simplified in the absence of energy and gap degeneracies.  Under this assumption, we have, combining Eqs.~\eqref{eq:LL3} and \eqref{eq:LL4}: 
\begin{align}
L_{\beta,\omega,ac}L_{\alpha,\omega,da}^{\dagger}  &= A_{\alpha,ca} A_{\beta,{ac}} \delta_{\omega, \eps_c - \eps_a} \delta_{dc}\ ,\\
L_{\alpha,\omega,ac}^{\dagger} L_{\beta,\omega,cd} &= A_{\alpha, ac} A_{\beta, ca} \delta_{\omega,\eps_a - \eps_c} \delta_{ad}\ ,\\
L_{\alpha,\omega,cd}^{\dagger} L_{\beta,\omega,da} &= A_{\alpha, ad} A_{\beta, da} \delta_{\omega, \eps_a - \eps_d}\delta_{ca}\ .
\end{align}  
Using these identities,  we can simplify Eq.~\eqref{eqt:rate} to: 
\bes
\begin{align}
&\bra{a} \mathcal{L}_{\textrm{WCL}}[\rho] \ket{a} = \sum_{c} \sum_{\alpha,\beta} \left[ \gamma_{\alpha \beta}(\eps_c - \eps_a) A_{\beta,ac}  A_{\alpha,ca} \rho_{cc} \phantom{\frac{1}{2}} \right. \nonumber \\
&\  \left. -\gamma_{\alpha \beta}( \eps_a - \eps_c)  A_{\alpha,ac} A_{\beta,ca} \rho_{aa}  \right] \\
& =  \sum_{c \neq a} \sum_{\alpha,\beta} \left[ \gamma_{\alpha \beta}(\eps_c - \eps_a) A_{\beta,ac}  A_{\alpha,ca} \rho_{cc} \right. \nonumber \\
& \qquad\qquad \left. - \gamma_{\alpha \beta}( \eps_a - \eps_c) A_{\alpha,ac} A_{\beta,ca} \rho_{aa}  \right] \ ,
\end{align}
\ees
where in the second equality we used the fact that the $c=a$ terms cancel. The dissipative dynamics of the population of state $a$ is now in the form of a rate equation, with the positive terms representing repopulation of the state $a$, while the negative terms represent depopulation of the state $a$.  Therefore we can identify the rate of loss of population from the $a$th energy eigenstate due to the dissipative dynamics as 
\beq
r_a = \sum_{\alpha, \beta} \sum_{c \neq a} \gamma_{\alpha \beta} (\eps_a - \eps_c)  A_{\alpha, a c}  A_{\beta, ca}   \ .
\eeq
We obtain Eq.~\eqref{eq:r0} upon restricting to the ground state ($a=0$) and assuming independent baths [$\gamma_{\alpha\beta}(\omega) = \delta_{\alpha\beta}\gamma(\omega)$].

\end{document}